\documentclass[showpacs,aps,twocolumn]{revtex4}
\usepackage{bm}
\usepackage{amsmath}
\usepackage{graphicx}
\usepackage{subfigure}
\usepackage[usenames,dvipsnames]{color}
\definecolor{darkblue}{RGB}{0,0,196}
\usepackage[colorlinks=true,linkcolor=darkblue,citecolor=darkblue,urlcolor=darkblue]{hyperref}
\usepackage{setspace}
\usepackage{hyperref}
\usepackage{footmisc}
\usepackage[makeroom]{cancel}
\usepackage{comment}
\usepackage{lineno}
\def\be{\begin{equation}}
\def\ee{\end{equation}}
\def\ba{\begin{eqnarray}}
\def\ea{\end{eqnarray}}
\usepackage{graphicx}
\usepackage{amsmath,bbm}
\usepackage{amssymb,bm}
\usepackage{yfonts}
\usepackage{adjustbox}
\usepackage{multirow}

\begin{document}
\title{Event shape and Multiplicity dependence of Freeze-out Scenario and System Thermodynamics in Proton+Proton Collisions at $\sqrt{s}$ = 13 TeV using PYTHIA8}
\author{Sushanta Tripathy}
\author{Ashish Bisht} 
\author{Raghunath Sahoo\footnote{Corresponding Author Email: Raghunath.Sahoo@cern.ch}}
\affiliation{Department of Physics, Indian Institute of Technology Indore, Simrol, Indore 453552, India}
\author{Arvind Khuntia}
\affiliation{Institute of Nuclear Physics, Polish Academy of Science (PAS), IFJ PAN, Krakow, Poland}
\author{Malavika P S}
\affiliation{National Institute of Technology Tiruchirappalli, Tiruchirappalli,  Tamil Nadu, India}
\begin{abstract}
\noindent
Recent observations of QGP-like conditions in high-multiplicity pp collisions from ALICE experiment at the LHC warrants an introspection whether to use pp collisions as a baseline measurement to characterize heavy-ion collisions for possible formation of a Quark-Gluon Plasma. A double differential study of the particle spectra and  thermodynamics of the produced system as a function of charged-particle multiplicity and transverse spherocity in pp collisions would shed light into the underlying event dynamics. Transverse spherocity, one of the event shape observables, allows to separate the events in terms of jetty and isotropic events. We analyse the identified particle transverse momentum ($p_{\rm T}$) spectra as a function of charged-particle multiplicity and transverse spherocity using Tsallis non-extensive statistics and Boltzmann-Gibbs Blastwave (BGBW) model in pp collisions at $\sqrt{s}$ = 13 TeV using PYTHIA8 event generator. The extracted parameters such as temperature ($T$), radial flow ($\beta$) and non-extensive parameter ($q$) are shown as a function of charged-particle multiplicity for different spherocity classes. We observe that the isotropic events approach to thermal equilibrium while the jetty ones remain far from equilibrium. We argue that, while studying the QGP-like conditions in small systems, one should separate the isotropic events from the spherocity-integrated events, as the production dynamics are different.
 
\pacs{13.85.Ni,12.38.Mh, 25.75.Nq, 25.75.Dw}
\end{abstract}

\date{\today}
\maketitle 

\section{Introduction}
\label{intro}
Although it was envisaged long back that central heavy-ion collisions at ultrarelativistic
energies could produce a deconfined state of partons called Quark-Gluon Plasma (QGP) \cite{Collins:1974ky,Cabibbo:1975ig}, the unprecedented collision energies available at the Large Hadron Collider (LHC) at CERN, Switzerland, has brought up new challenges in characterizing the proton+proton (pp) collisions 
to understand a possible formation of QGP droplets in these hadronic collisions. There are various signatures of QGP, which are already observed in pp collisions at the LHC. These include, strangeness enhancement \cite{ALICE:2017jyt}, hardening of $p_{\rm T}$-spectra~\cite{Tripathy:2018ehz,Dash:2018cjh} and the thermal effective temperature being comparable to that observed in heavy-ion collisions \cite{ALICE:2017jyt}, degree of collectivity \cite{Khuntia:2018znt}  etc. In view of these observations in pp collisions, it has become more 
challenging to understand the system formed in pp collisions, although pp has been considered as a baseline measurement to understand nuclear effects like $R_{\rm AA}$, suppression of J/$\psi$ etc. The new measurements at the LHC keeping in mind that the final state multiplicity drives the particle production (excellent scaling observed), necessitates  a closer look into the underlying physics mechanisms in particle production in pp collisions. The phenomena like color reconnection, multipartonic interactions, rope hadronization, string fragmentation etc. have done a wonderful job in explaining various new heavy-ion-like observations in pp collisions.

Event shape engineering has given a new direction to underlying events in pp collisions to have a differential study taking various observables. The transverse spherocity successfully separates jetty events from isotropic ones in pp collisions. As is clearly understandable, the particle production mechanism in jetty events are different from isotropic ones. When the former one involves high-$p_{\rm T}$ phenomena, the latter is soft-physics dominated. In view of this, recently we have carried out a double differential study of particle ratios in pp collisions at the LHC energies, taking transverse spherocity, transverse momentum and 
multiplicity \cite{Khuntia:2018qox}. A natural question which pops up is- whether the thermodynamics of jetty events are different from the isotropic ones. To quantify this, in the present study, we have taken pQCD-inspired PYTHIA8 event generator which includes multi-parton interactions (MPI) along with color reconnection (CR), to study the event shape and multiplicity dependence of freeze-out scenario and system thermodynamics in pp collisions at $\sqrt{s}$ = 13 TeV. It has been reported that, MPI scenario is crucial to explain the underlying events, multiplicity distributions and flow-like patterns in terms of color reconnection~\cite{Ortiz:2013yxa}. Thus, it is a preferable tune to study the possible thermodynamics in small systems, as experimental data are not available yet. 
It should be worth noting that PYTHIA8 does not have inbuilt thermalization. However, as reported in Ref.~\cite{Ortiz:2013yxa}, the color reconnection (CR) mechanism along with the multi-partonic interactions (MPI) in PYTHIA8 produces the properties which arise from thermalization of a system such as radial flow and mass dependent rise of mean transverse momentum. In PYTHIA model, a single string connecting two partons follows the movement of the partonic endpoints and this movement gives a common boost to the string fragments (final state hadrons). With CR along with MPI, two partons from independent hard scatterings can reconnect and they increase the transverse boost. This microscopic treatment of final state particle production is quite successful in explaining the similar features which arise from a macroscopic picture via hydrodynamical description of high-energy collisions. Thus, it is apparent to say that the PYTHIA8 model with MPI and CR, has plausible ability to produce the features of thermalization.
The current results, which use PYTHIA8 with MPI and CR to obtain event shape and multiplicity dependence of freeze-out scenario and system thermodynamics, will help to compare with the upcoming experimental data. Such a study has also been done for heavy flavor particles like J/$\psi$ in Ref.~\cite{Khatun:2019dml}. This paper is intended solely for presenting a noble and unique study, which would give an outlook on similarities/differences between jetty and isotropic events in LHC pp events and their multiplicity dependence. This will help in making a
proper bridge in understanding the particle production from hadronic to heavy-ion collisions. 

Further, the spacetime evolution of hadronic and heavy-ion collisions at the LHC energies could be thought of following a cosmological expansion of the
produced fireball. In this scenario, as the fireball expands and cools down, it leaves a temperature profile with time. Different 
identified particles decouple from the fireball giving the signature of a mass dependent particle freeze-out -- higher mass particles decoupling from the system earlier in time. In this work, we have considered such a scenario and have performed a differential study
taking final state event multiplicity and event topology.

The paper is organized as follows. After the introduction and identifying of the problem under consideration, we discuss the methodology of event generation and data analysis in Section-\ref{ev_ana}. In Section-\ref{sec-III}, we discuss the identified $p_{\rm T}$-spectra in pp collisions to extract the thermodynamic parameters. Finally we summarize the work in Section-\ref{summary} with important findings, which could be tested when experimental data become available.

\begin{figure}[ht]
\includegraphics[width=8.5cm, height=7.0cm]{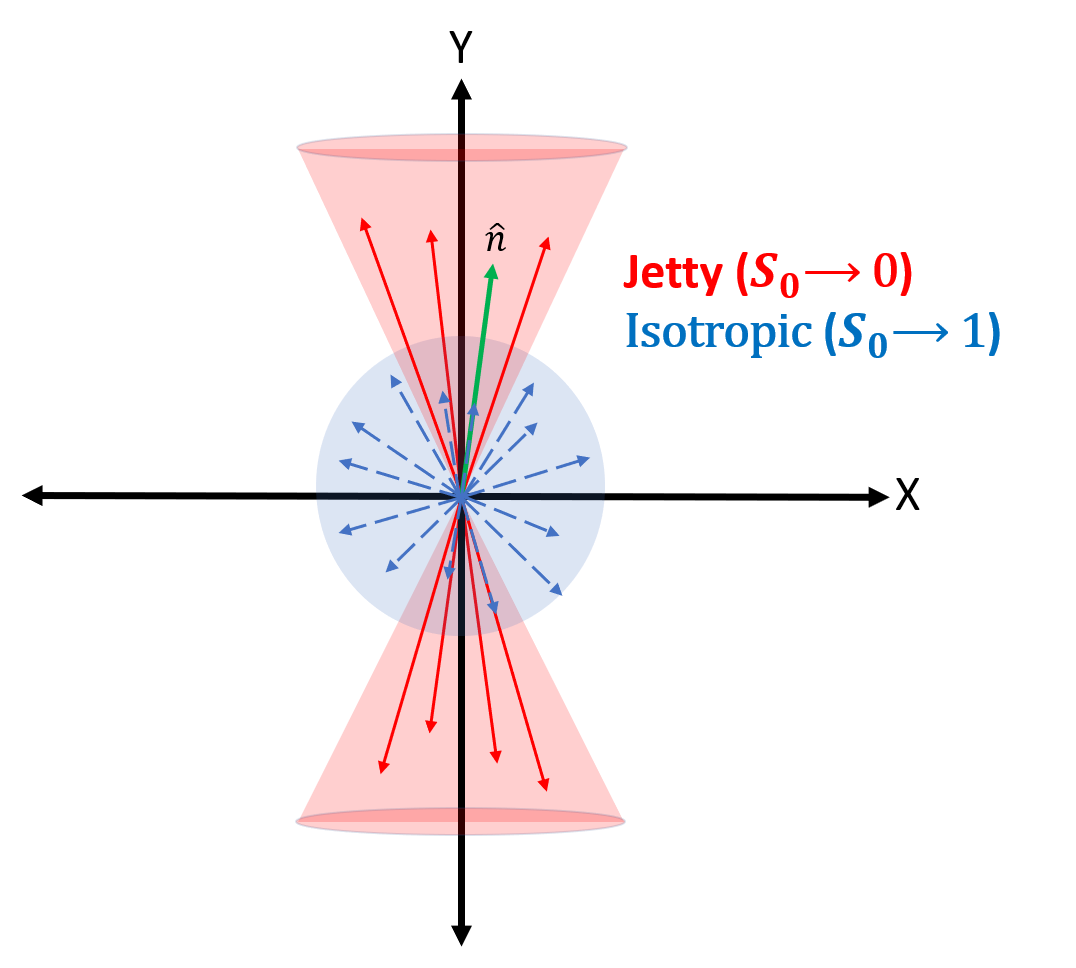}
\caption[]{(Color Online) Figure showing jetty and isotropic events in the transverse plane.}
\label{sp_cart}
\end{figure}

\begin{figure}[ht!]
\includegraphics[scale=0.4]{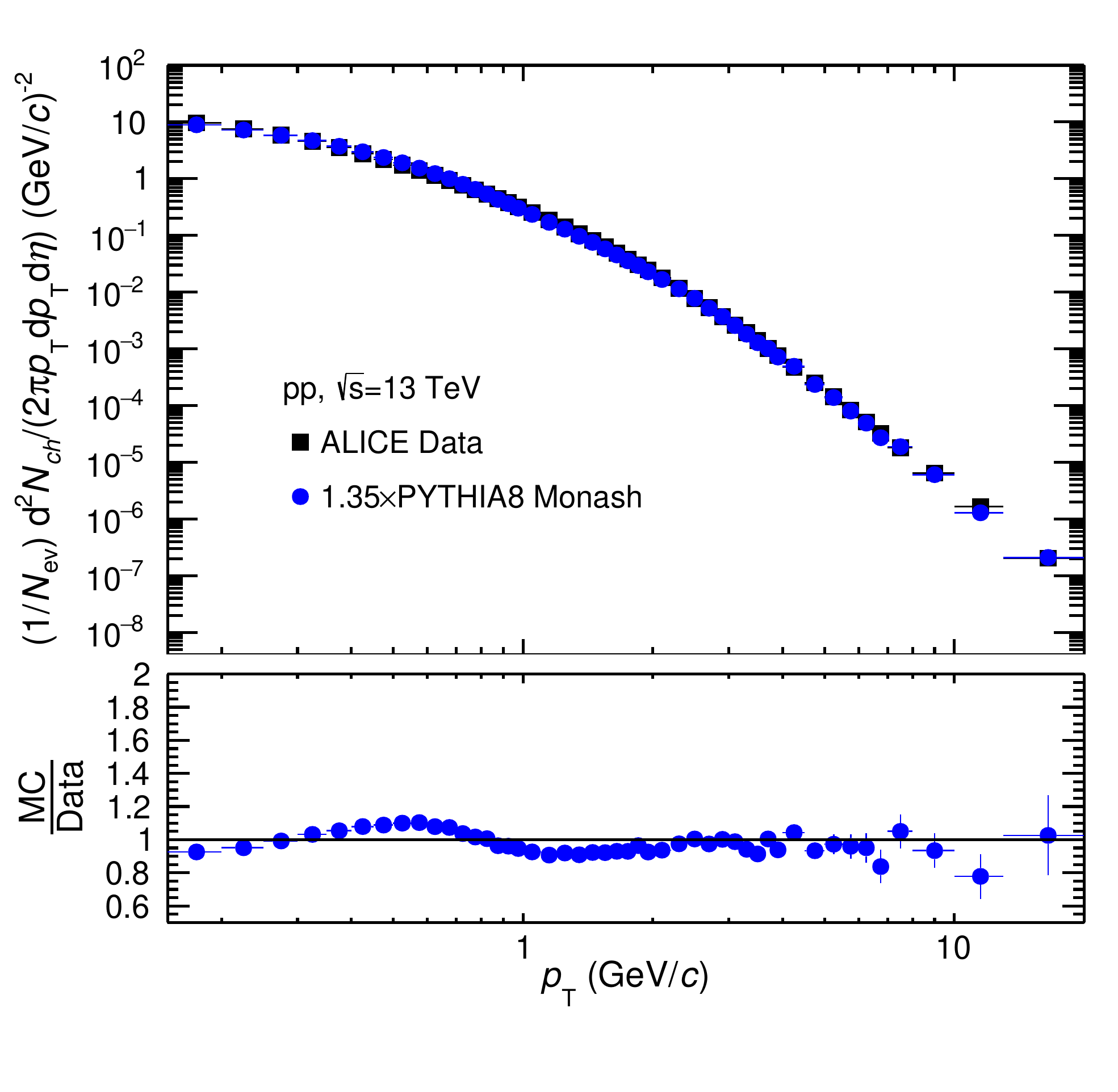}
\caption[]{(Color Online) Upper Panel: Comparison of charged particle $p_{\rm T}$ spectra in pp collisions at $\sqrt s$ = 13 TeV between ALICE data \cite{Adam:2015pza} and PYTHIA8 simulation, which is used for this analysis. Lower Panel: the ratio between scaled simulated data and experimental data.}
\label{datVsPH}
\end{figure}

\section{Event Generation and Analysis Methodology}
\label{ev_ana}
PYTHIA, one of the popular and most useful event generators in the LHC era, is used to simulate ultra-relativistic collision events among the elementary particles like $e^{\pm}, $~p$,\rm{~and~} \bar{p} $. It is incorporated with many known physics mechanisms like hard and soft interactions, parton distributions, initial- and final-state parton showers, multipartonic interactions, string fragmentation, color reconnection and resonance decays~\cite{Sjostrand:2006za}.

In our present study, we have used PYTHIA 8.235 to generate pp collisions at $\sqrt{s}$ = 13 TeV with Monash 2013 Tune (Tune:14)~\cite{Skands:2014pea}. PYTHIA 8.235 is an advanced version of PYTHIA 6 which includes the multi-partonic interaction (MPI) scenario as one of the key improvements. The detailed physics processes in PYTHIA 8.235 can be found in Ref.~\cite{pythia8html}. We have implemented the inelastic, non-diffractive component of the total cross-section for all soft QCD processes with the switch SoftQCD : all = on. This analysis is carried out with around 250 million minimum bias events at $ \sqrt{s}=13~\mathrm{TeV}$ and we have chosen MPI based scheme of  default color reconnection mode (ColorReconnection:mode(0)). Here, the minimum bias events are those events where no selection on charged particle multiplicity and/or spherocity is applied. For the generated events we let all the resonances to decay except the ones used in our study with the switch HadronLevel:Decay = on. Throughout the analysis, the event selection criteria is such that only those events were chosen which have at-least 5 charged particles. To match with experimental conditions, charged particle multiplicities ($N_{\rm ch}$) have been chosen in the acceptance of V0 detector in ALICE at the LHC with pseudorapidity coverage of V0A ($2.8<\eta<5.1$) and V0C ($-3.7<\eta<-1.7$)~\cite{Abelev:2014ffa}. The generated events are categorized in ten V0 multiplicity (V0M) bins, each with 10\% of the total number of events. The number of charged particle multiplicities in an event in different V0 multiplicity classes are listed in Table~\ref{tab:V0M}. 

\begin{table}[h]
\caption{V0 multiplicity classes and the corresponding charged particle multiplicities.}
\centering 
\scalebox{0.8}
{
\begin{tabular}{|c|c|c|c|c|c|c|c|c|c|c|} 
\hline % inserting double-line
V0M class & I & II & III & IV & V & VI & VII &VIII & IX & X \\
\hline 
$N_{ch}$ &50-140 & 42-49 & 36-41 & 31-35 & 27-30 & 23-26 & 19-22 &  15-18 & 10-14 & 0-9\\
\hline
\end{tabular}
}
\label{tab:V0M}
\end{table}

For an event, transverse spherocity is defined for a unit vector $\hat{n} (n_{T},0)$ which minimizes the ratio~\cite{Cuautle:2014yda, Cuautle:2015kra, Ortiz:2017jho}:
\begin{eqnarray}
S_{0} = \frac{\pi^{2}}{4} \bigg(\frac{\Sigma_{i}~|\vec p_{T_{i}}\times\hat{n}|}{\Sigma_{i}~p_{T_{i}}}\bigg)^{2}.
\end{eqnarray}

By restricting it to transverse plane, transverse spherocity becomes infrared and collinear safe~\cite{Salam:2009jx} and by construction, the extreme limits of transverse spherocity are related to specific configurations of events in transverse plane. The value of transverse spherocity ranges from 0 to 1. Transverse spherocity becoming 0 means that the events are pencil-like (back to back structure) while 1 would mean the events are isotropic as shown in Fig.~\ref{sp_cart}. The pencil-like events are usually the hard events while the isotropic events are the result of soft processes. Here onwards, for the sake of simplicity the transverse spherocity is referred as spherocity. To disentangle the jetty and isotropic events from the average-shaped events, we have applied spherocity cuts on our generated events. In this analysis, the spherocity distributions are selected in the pseudo-rapidity range of $|\eta|<0.8$ with a minimum constraint of 5 charged particles with $p_{\rm{T}}~$ $>$ 0.15 GeV/$c$. The jetty events are those having $0\leq S_{0}<0.29$ with lowest 20 percent ($\simeq$ 50M events) and the isotropic events are those having $0.64<S_{0}\leq1$ with highest 20 percent ($\simeq$ 50M events) of the total events~\cite{Bencedi:2018ctm}.

\begin{figure}[ht!]
\includegraphics[width=8.cm, height=7.cm]{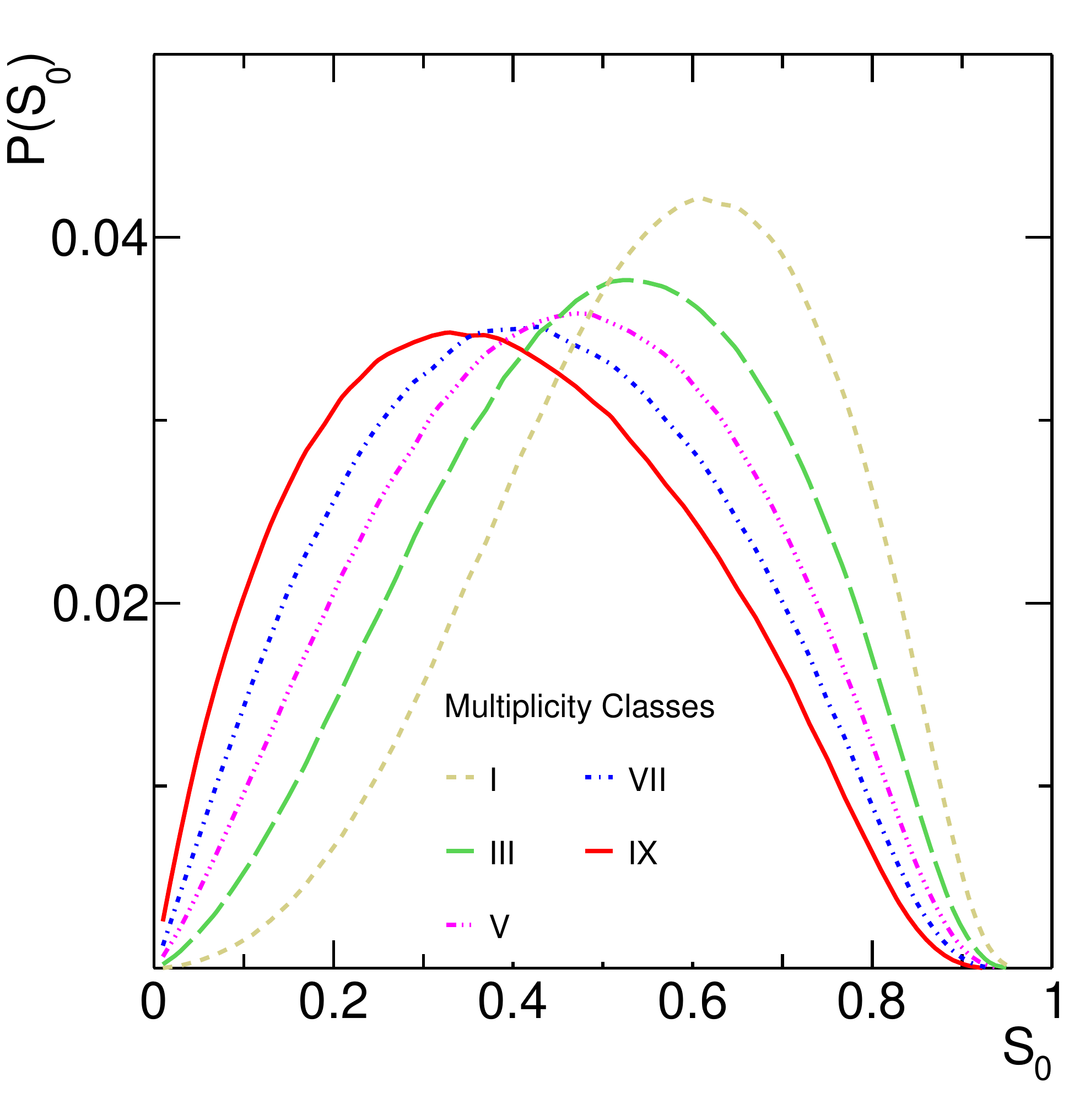}
\caption[]{(Color Online) Spherocity distribution for different multiplicity classes in pp collisions at $ \sqrt{s} =\mathrm{13~TeV}$ using PYTHIA8. Different colors and line styles are for different multiplicity classes. }
\label{sp_dis}
\end{figure}

To assure the quality of the generated events, we show in Fig.~\ref{sp_dis}, the correlation between the spherocity with charged particle multiplicity. As expected, the high multiplicity pp collisions are dominated by isotropic events while the low multiplicity events are dominated by the jetty ones. From our earlier event shape analysis~\cite{Khuntia:2018qox}, it is evident that spherocity along with the charged particle multiplicity (which is correlated with nMPI) should be preferred for a better selectivity of events. 
%\clearpage
\section{Transverse momentum spectra of identified particles}
\label{sec-III}
To check the compatibility of PYTHIA8 simulated data with the experimental data, we have compared the charged particle $p_{\rm T}$ spectra for pp collisions at $\sqrt{s}$ = 13 TeV from ALICE data ~\cite{Adam:2015pza}. The comparison is shown in Fig. \ref{datVsPH}. The lower panel show the ratio of the predictions from PYTHIA8 to experimental data. In order to see the agreement of spectral shapes, we have used arbitrary scaling factor (1.35) to scale the simulated data. The used scaling factor is to check the matching of the spectral shape and it bears no physical significance. We found that the scaled simulated data agree with the spectral shape from experimental data within (10-20)\% at low-$p_{\rm T}$ and consistent to unity for intermediate and high-$p_{\rm T}$.

For the first time, we combine spherocity with event multiplicity and study the freeze-out scenario and thermodynamics of the system formed in pp collisions at $\sqrt{s}$ = 13 TeV. We use experimentally 
motivated thermodynamically consistent Tsallis non-extensive distribution function~\cite{Cleymans:2011in} for 
analysing the complete range of the $p_{\rm T}$-spectra, whereas to extract the kinetic freeze-out temperature
and the possible collective radial flow we use Boltzmann-Gibbs Blastwave model~\cite{Cooper,Schnedermann:1993ws} taking $p_{\rm T} \leq$ 2 GeV/c. We begin with the fitting and analysis procedure with short description on Tsallis non-extensive statistics and Boltzmann-Gibbs Blast wave model. Here onwards, $(\pi^{+}+\pi^{-}), (\rm{K}^{+}+\rm{K}^{-}),~(p+\bar{p}),~(\rm{K}^{*0}+\overline{\rm{K}^{*0}})~\rm{and} ~(\Lambda+\bar{\Lambda})$ are denoted as pion ($\pi$), kaon (K), proton (p), $\rm{K}^{*0}$ and $\Lambda$, respectively.

\subsection{Experimentally motivated Tsallis non-extensive statistics}
\label{tsallis}
The $p_{\rm{T}}$-spectra of produced particles in high-energy collisions have been proposed to follow a thermalised Boltzmann type of distribution given as~\cite{Hagedorn:1965st},

\begin{eqnarray}
\label{eq2}
E\frac{d^3\sigma}{d^3p}& \simeq C \exp\left(-\frac{p_T}{T_{kin}}\right).
\end{eqnarray}
Here $C$ is the normalisation constant and $T_{\rm kin}$ is the kinetic freeze-out temperature. Due to possible QCD contributions at high-$p_{\rm T}$, the identified particle spectra at RHIC and LHC do not follow the above distribution, while the low-$p_{\rm T}$-region can be explained by incorporating the radial flow ($\beta$) into Boltzmann-Gibbs distribution function, which is known as Boltzmann-Gibbs Blast Wave (BGBW) model~\cite{Schnedermann:1993ws}. One can extract $T_{\rm kin}$ and radial flow ($\beta$) by fitting the identified particle transverse momentum spectra at low-$p_{\rm T}$. The detailed description along with the fitting of Boltzmann-Gibbs Blast Wave model to the identified particle spectra is discussed in the next sub-section.      

\begin{figure}[ht!]
\includegraphics[width=9cm, height=11.cm]{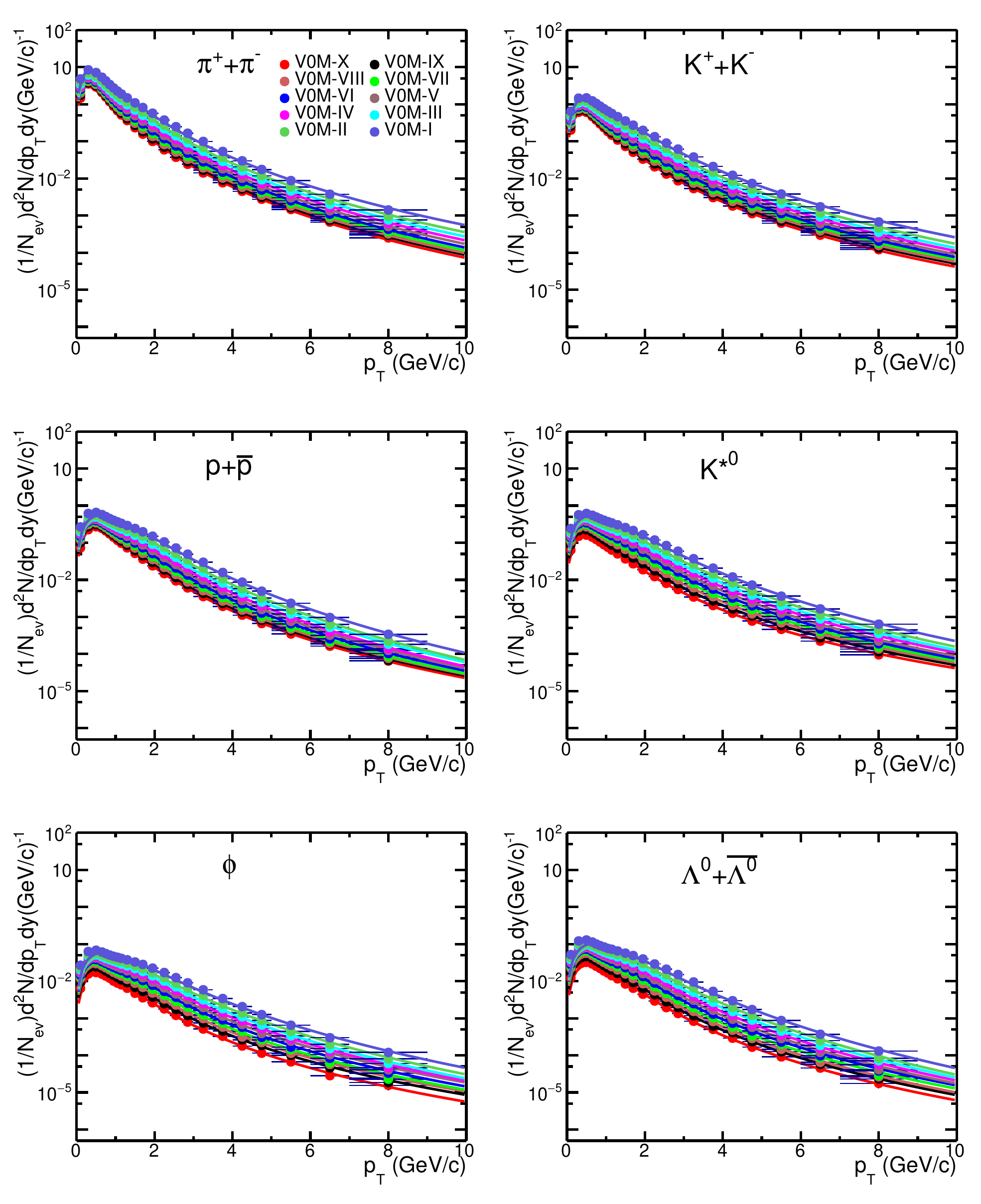}
\caption[]{(Color Online) Fitting of generated $p_{\rm{T}}$-spectra of identified hadrons from PYTHIA8 using Tsallis distribution for spherocity integrated events in various multiplicity classes as shown in Table~\ref{tab:V0M}.}
\label{tsallis_S0}
\end{figure}

\begin{figure}[ht!]
\includegraphics[width=9cm, height=11.cm]{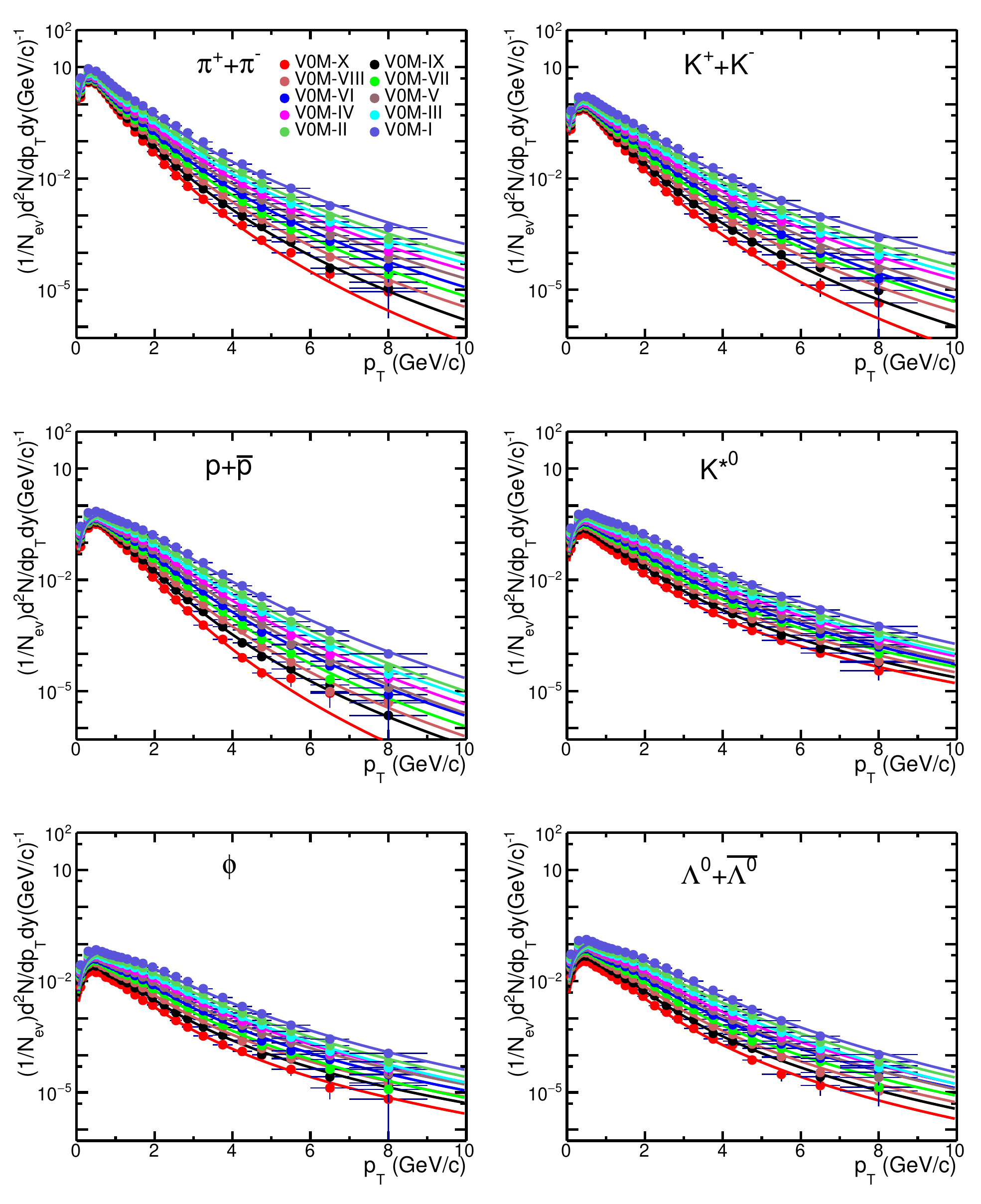}
\caption[]{(Color Online) Fitting of generated $p_{\rm{T}}$-spectra of identified hadrons from PYTHIA8 using Tsallis distribution for isotropic events in various multiplicity classes as shown in Table~\ref{tab:V0M}.}
\label{tsallis_iso}
\end{figure}

\begin{figure}[ht!]
\includegraphics[width=9cm, height=11.cm]{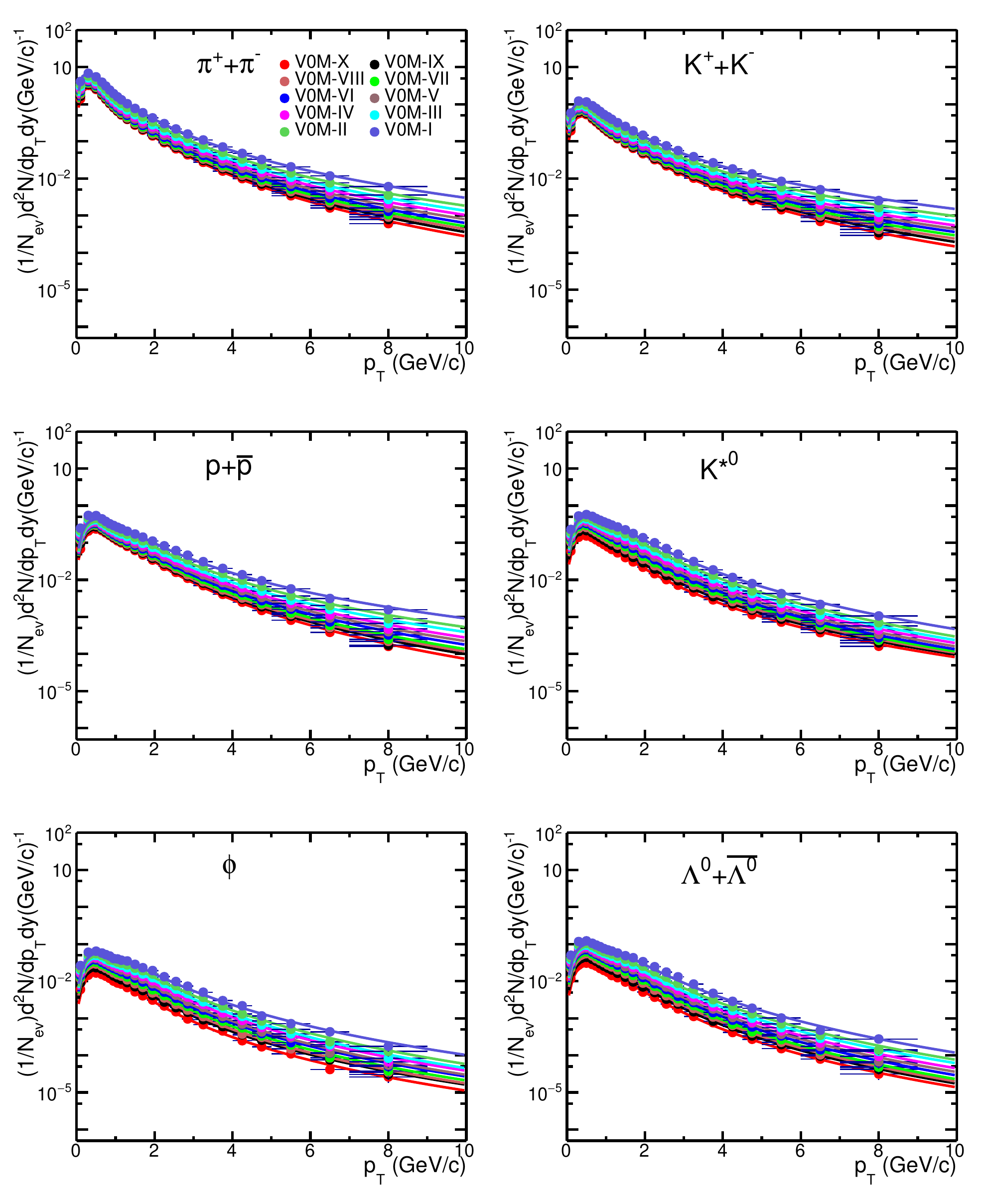}
\caption[]{(Color Online) Fitting of generated $p_{\rm{T}}$-spectra of identified hadrons from PYTHIA8 using Tsallis distribution for jetty events in various multiplicity classes as shown in Table~\ref{tab:V0M}.}
\label{tsallis_jetty}
\end{figure}

To describe the complete $p_{\rm T}$-spectra, one has to account for the power-law contribution at high-$p_{\rm T}$  \cite{CM,CM1,UA1}, which empirically takes care of the possible QCD contributions. A combination of both low and high-$p_{\rm T}$ aspects has been proposed by Hagedorn, which describes the experimental data over a wide  $p_{\rm T}$-range~\cite{Hagedorn:1983wk}. The distribution proposed by Hagedron is given by 
  
\begin{eqnarray}
  E\frac{d^3\sigma}{d^3p}& = &C\left( 1 + \frac{p_T}{p_0}\right)^{-n}
\nonumber\\
 & \longrightarrow&
  \left\{
 \begin{array}{l}
  \exp\left(-\frac{n p_T}{p_0}\right)\quad \, \, \, {\rm for}\ p_{\rm T} \to 0, \smallskip\\
  \left(\frac{p_0}{p_T}\right)^{n}\qquad \qquad{\rm for}\ p_{\rm T} \to \infty.
 \end{array}
 \right .
 \label{eq3}
\end{eqnarray}
Here, $C$, $p_0$, and $n$ are fitting parameters. The above expression acts as an exponential and a power-law function for low and high-$p_{\rm T}$, respectively. However, deviations are observed by experiments at RHIC~\cite{star-prc75,phenix-prc83} and LHC~\cite{alice1,alice2,alice3,cms} while describing the $p_{\rm T}$-spectra of identified particles using a Boltzmann-Gibbs distribution function, even if the domain of temperature
of the produced systems are high enough. On the other hand, Tsallis statistics with its non-extensivity features can be regarded as a generalization of Boltzmann-Gibbs statistics and it gives a better description of systems which have not yet reached equilibration. Its low-$p_{\rm T}$ exponential and high-$p_{\rm T}$ power-law behaviour gives a complete spectral description of identified secondaries produced in pp collisions. In addition, a non-extensive entropic $q$-parameter shows the extent of non-equilibration of any particle in a thermal bath.  There are few different versions of Tsallis distribution, which are being used by experimentalists and theoreticians. However, we use a thermodynamically consistent Tsallis non-extensive distribution function as shown in Ref.~\cite{Cleymans:2011in}. By saying thermodynamically consistent, we mean that the used distribution function satisfies all the standard thermodynamic relations for entropy, temperature, energy, pressure and number density.  The Tsallis distribution function at mid-rapidity is given by,

\begin{eqnarray}
\label{eq4}
\left.\frac{1}{p_T}\frac{d^2N}{dp_Tdy}\right|_{y=0} = \frac{gVm_T}{(2\pi)^2}
\left[1+{(q-1)}{\frac{m_T}{T}}\right]^{-\frac{q}{q-1}}.
\end{eqnarray}
 where, $g$ is the degeneracy factor, $V$ is the system volume, $m_{\rm T} = \sqrt{p_T^2 + m^2}$ is the transverse mass and $q$ is the non-extensive parameter. In the limit of $q \rightarrow 1$, Tsallis distribution (Eq.~\ref{eq4}) reduces to the standard Boltzmann-Gibbs distribution (Eq.~\ref{eq2}). 
 It should be noted here that the use of Tsallis non-extensive distribution function is purely motivated by its excellent description of experimental particle spectra. It has been widely used to explain the particle spectra in high-energy collisions~\cite{Bhattacharyya:2015nwa,Bhattacharyya:2015hya,Zheng:2015gaa,Tang:2008ud,De:2014dna} starting from elementary $e^++e^-$ and hadronic to heavy-ion collisions~\cite{e+e-,R1,R2,R3,ijmpa,plbwilk,marques,STAR,PHENIX1,PHENIX2,ALICE_charged,ALICE_piplus,CMS1,CMS2,ATLAS,ALICE_PbPb}. Recently, few comprehensive studies have been carried out using Tsallis distribution for pions and quarkonium spectra in pp collisions~\cite{Grigoryan:2017gcg,Parvan:2016rln}. In this subsection, we employ the  Tsallis non-extensive distribution function as shown in Eq.~\ref{eq4} to describe the $p_{\rm{T}}$-spectra. It should be noted here that the used Tsallis distribution function of Eq.~\ref{eq4} \cite{Cleymans:2013rfq} has an extra power $q$, compared to the original distribution function proposed by 
 C. Tsallis \cite{book}. However, this form of the distribution function is
  thermodynamically consistent, which makes no major change in the observables as the values of $q$ in hadronic collisions lie
  between, $1 \leq q \leq 1.22$ \cite{{Beck:2003kz}}. While Tsallis non-extensive statistics makes a connection between entropy to thermodynamics of a system, the dynamics of the system in terms of long range correlations and fluctuations ($(q-1)$ being the strength of the fluctuation \cite{Wilk}) are encoded in the entropic parameter, $q$.

\begin{figure*}[ht!]
\includegraphics[width=16cm, height=10.cm]{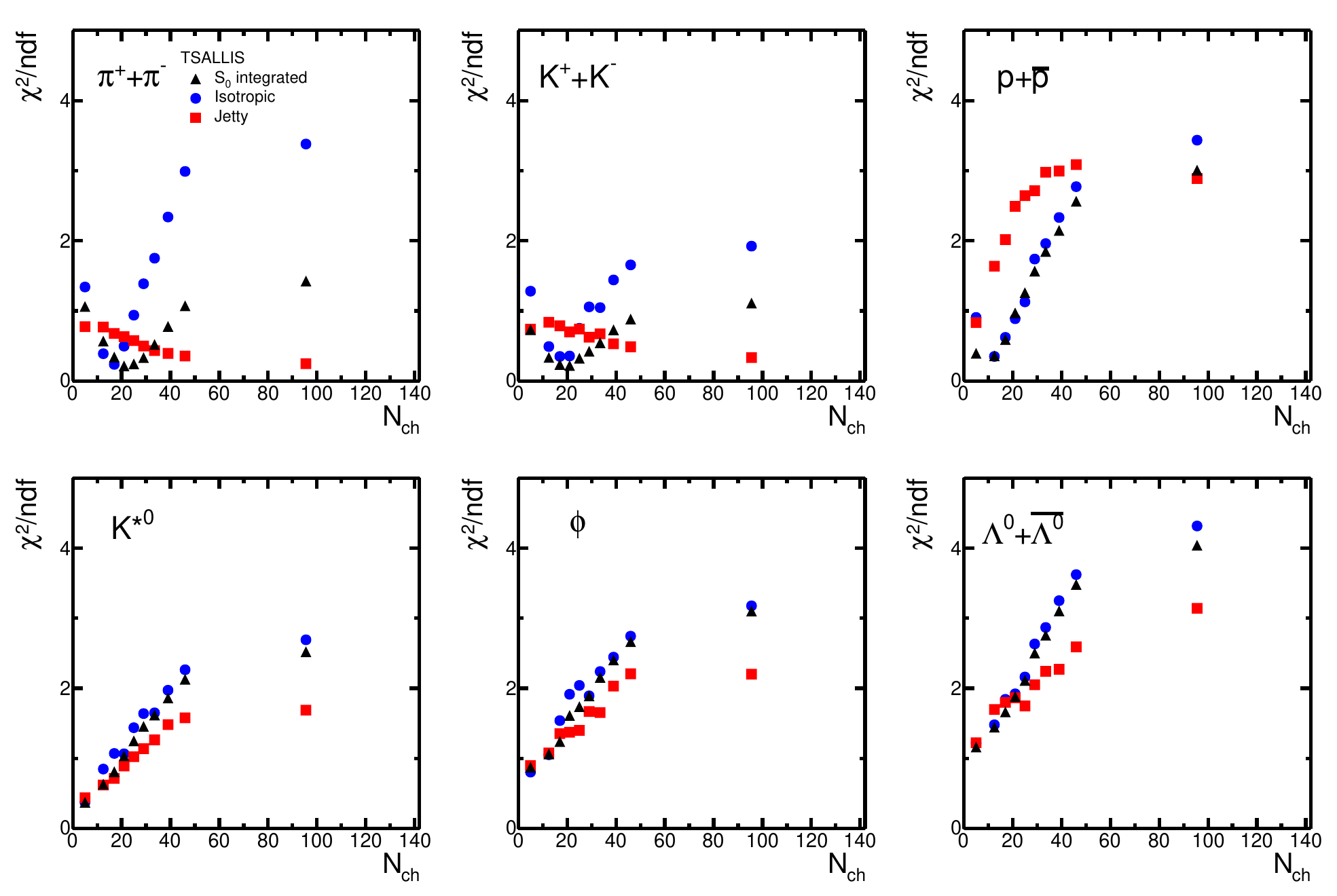}
\caption[]{(Color Online) $\chi^{2}$/NDF for the fitting of generated $p_{\rm{T}}$-spectra of identified hadrons using Tsallis distribution in different spherocity and multiplicity classes.}
\label{tsallis_chi2}
\end{figure*}

Figures~\ref{tsallis_S0},~\ref{tsallis_iso} and~\ref{tsallis_jetty} show the fitting of $p_{\rm{T}}$-spectra of pions, kaons, protons, $\rm{K}^{*0}$, $\phi$ and $\Lambda$ as a function of charged-particle multiplicity using Tsallis distribution function (Eq.~\ref{eq4}) for spherocity-integrated, isotropic and jetty events, respectively. We observe that the Tsallis distribution fits the generated data till $p_{\rm{T}}\simeq$ 10 GeV/c. Figure~\ref{tsallis_chi2} shows the quality of fitting in terms of the reduced-$\chi^2$, $\chi^2/NDF$ as a function of multiplicity for different spherocity classes. The values of $\chi^2/NDF$ shows that the quality of fitting is reasonably good for all the multiplicity and spherocity classes.

\begin{figure*}[ht!]
\includegraphics[width=16cm, height=10.cm]{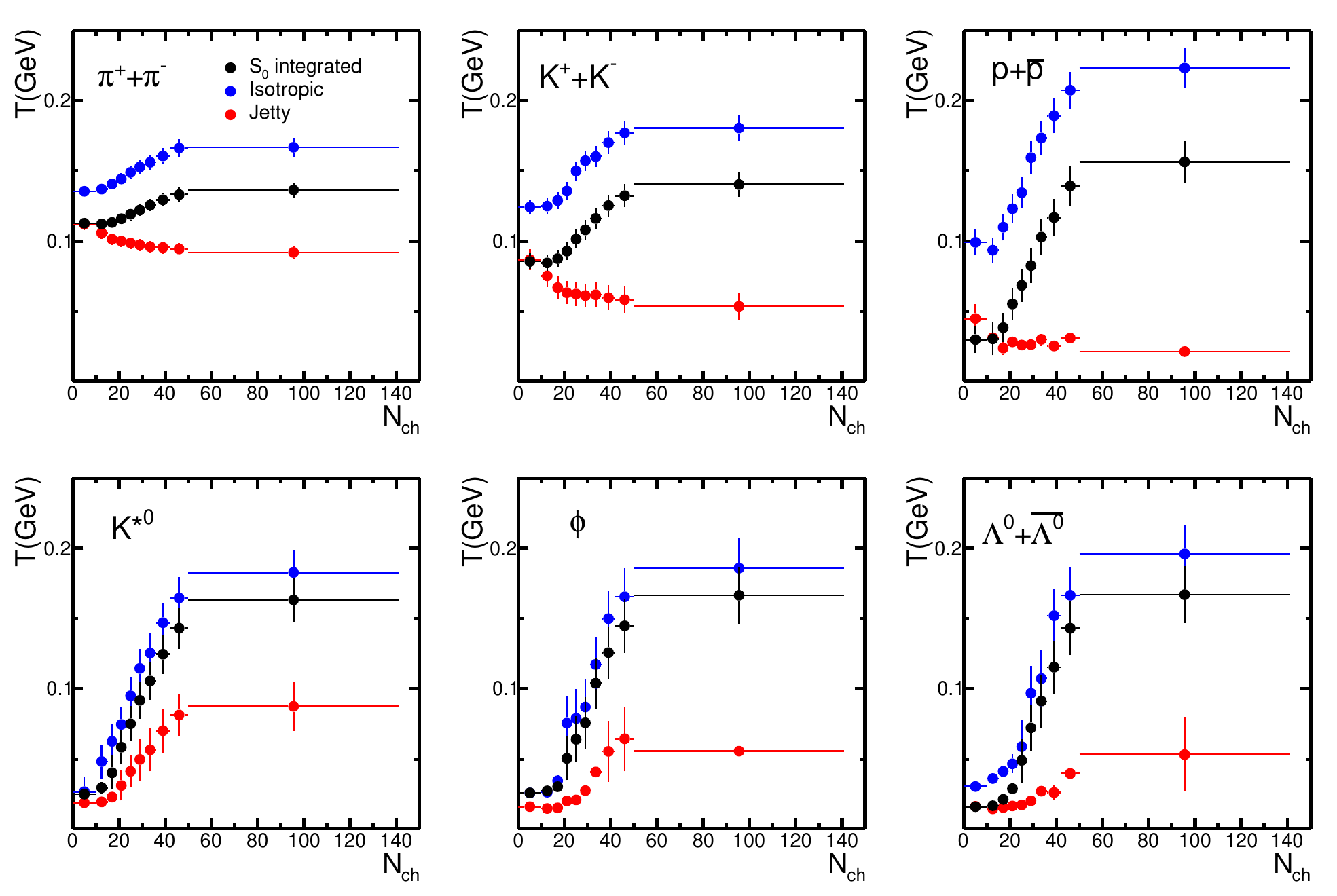}
\caption[]{(Color Online) Multiplicity dependence of $T$ in different spherocity classes from the fitting of Tsallis distribution using Eq.~\ref{eq4} }
\label{tsallis_param_T}
\end{figure*}

\begin{figure*}[ht!]
\includegraphics[width=16cm, height=10.cm]{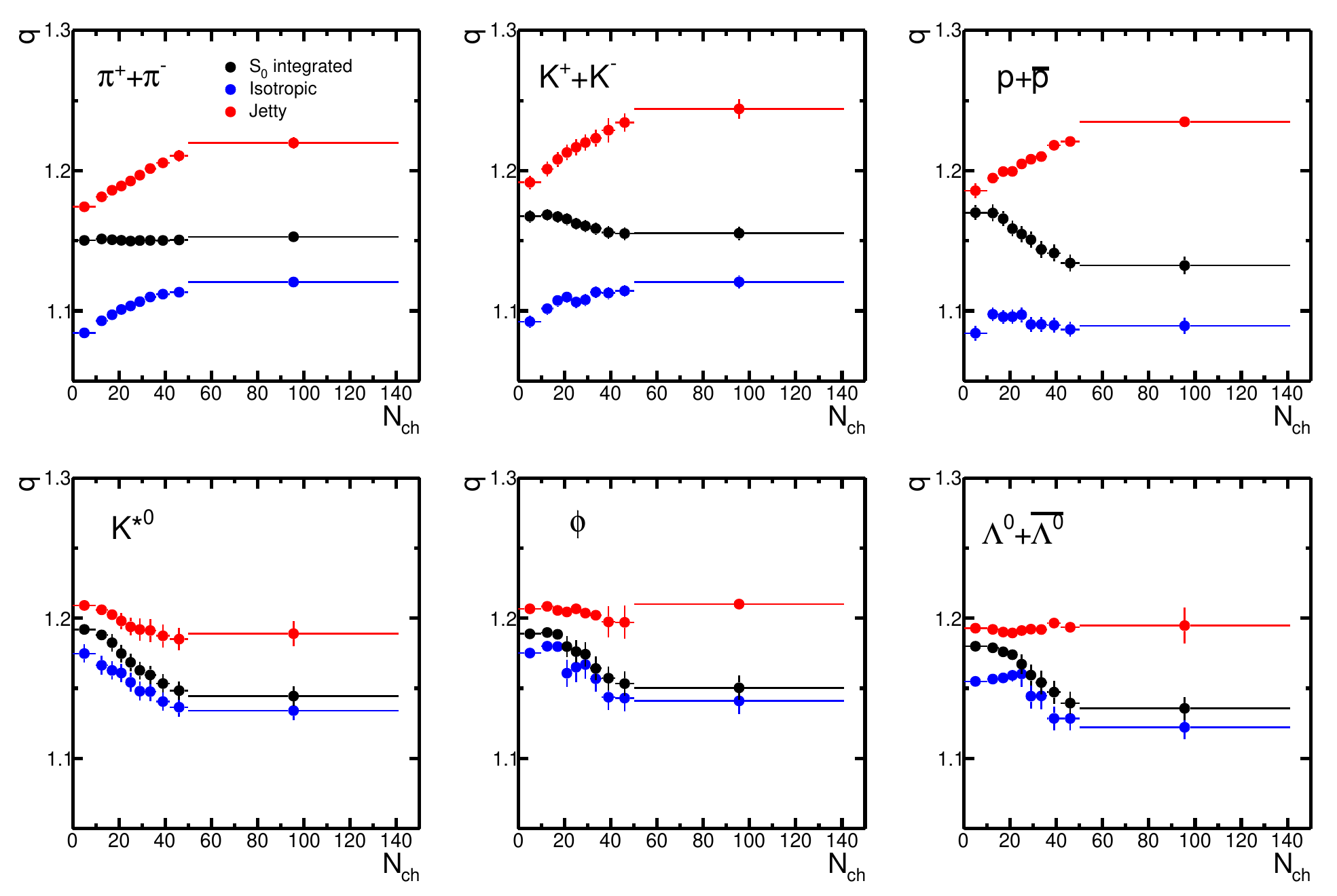}
\caption[]{(Color Online) Multiplicity dependence of $q$ in different spherocity classes from the fitting of Tsallis distribution using Eq.~\ref{eq4} }
\label{tsallis_param_q}
\end{figure*}

\begin{figure}[ht!]
\includegraphics[width=8cm, height=12.cm]{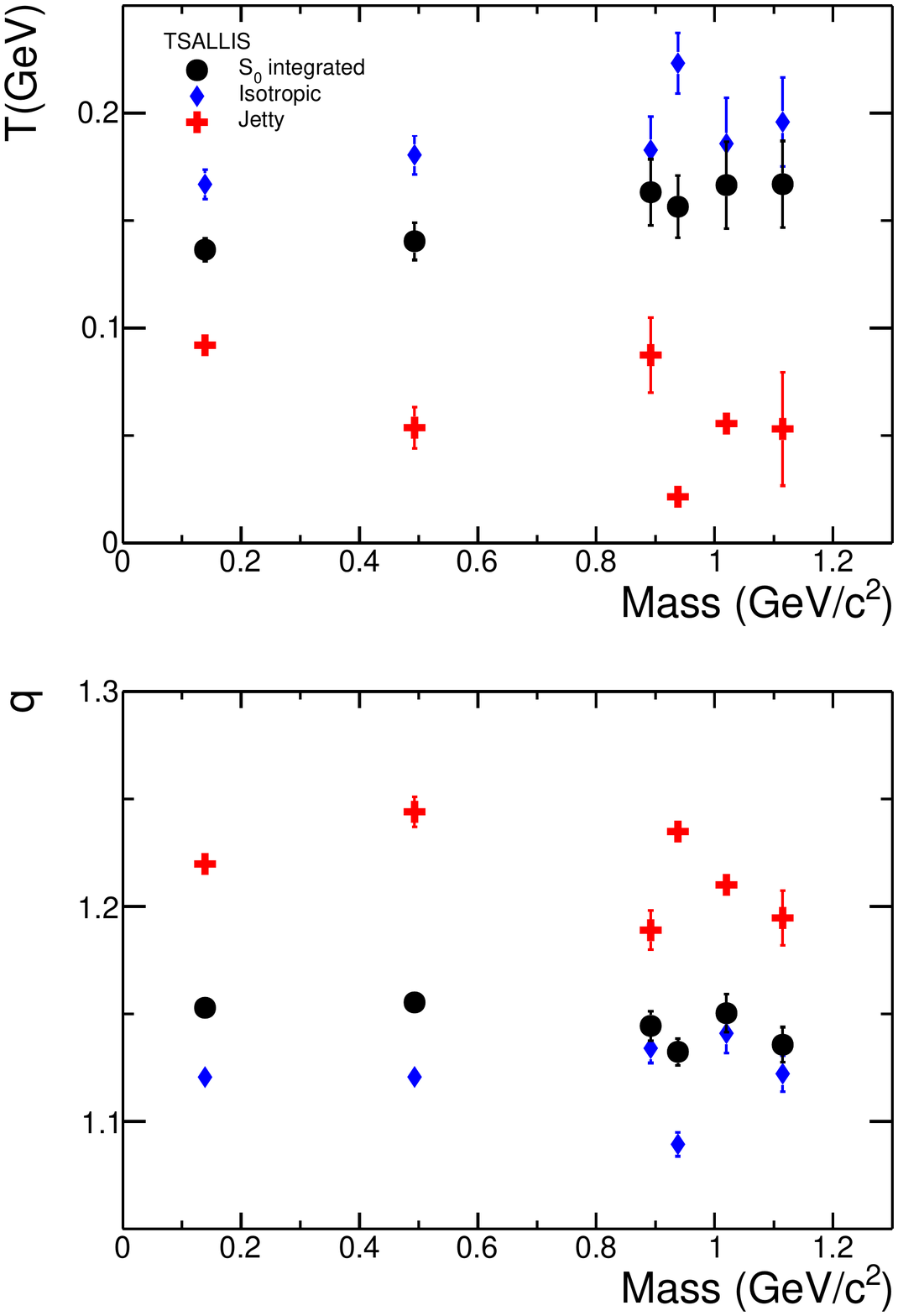}
\caption[]{(Color Online) Mass dependence of T and $q$ in different spherocity classes for the highest multiplicity class.}
\label{tsallis_param_mass}
\end{figure}

Figures~\ref{tsallis_param_T} and~\ref{tsallis_param_q} show the extracted parameters from the fitting of Tsallis distribution using Eq.~\ref{eq4} as a function of charged-particle multiplicity for different spherocity classes. Figure~\ref{tsallis_param_T} shows that the temperature parameter increases with charged-particle multiplicity for spherocity-integrated and isotropic events. However, the jetty events seem to show a reverse trend for pions, kaons and protons. For $\rm{K}^{*0}$, $\phi$ and $\Lambda$ the temperature parameter shows an
increase with multiplicity for jetty events. For all cases, the temperature for jetty events is lower compared to the
other spherocity classes. We also observe that the temperature for lighter particles does not change significantly with multiplicity while with increase in mass, the temperature increases steeply as a function of multiplicity.
Figure~\ref{tsallis_param_q} shows that for isotropic events, the non-extensive parameter, $q$ values remain lower compared to the spherocity-integrated events which suggests that isotropic events have got a higher degree of equilibration compared to spherocity-integrated events. This indicates that while studying the QGP-like conditions in small systems, one should separate the isotropic events from the spherocity-integrated events, as the production dynamics are different. On the contrary, the $q$ value for jetty are always higher compared to spherocity-integrated events indicating that the jetty events remain far away from equilibrium. The present study is very useful in understanding the microscopic features of degrees of equilibration and their dependencies on the number of particles in the system and on the geometrical shape of an event. 
It would be interesting to study the particle mass dependence of these thermodynamic parameters. In order to
do that, we have taken the events with highest multiplicity class and done the same spherocity analysis taking different particles as discussed here, which is shown in Fig. \ref{tsallis_param_mass}. For the isotropic and spherocity-integrated events in high multiplicity pp collisions, the temperature remains higher for particles with higher masses which supports a differential freeze-out scenario. This suggests that massive particles freeze-out early from the system. However, the jetty events show a reverse trend. 

To explore the flow-like features in small systems, one needs to focus on the low-$p_{\rm{T}}$ of the particle spectra with Boltzmann-Gibbs Blastwave (BGBW) model, which is discussed in next sub-section. As we saw an indication
of a differential freeze-out scenario, in the following section we consider making individual spectral analysis using BGBW, instead of a simultaneous fitting, which is usually necessitated by a single freeze-out scenario. 

\subsection{Boltzmann-Gibbs Blastwave Model}
\begin{figure}[ht!]
\includegraphics[width=9cm, height=11.cm]{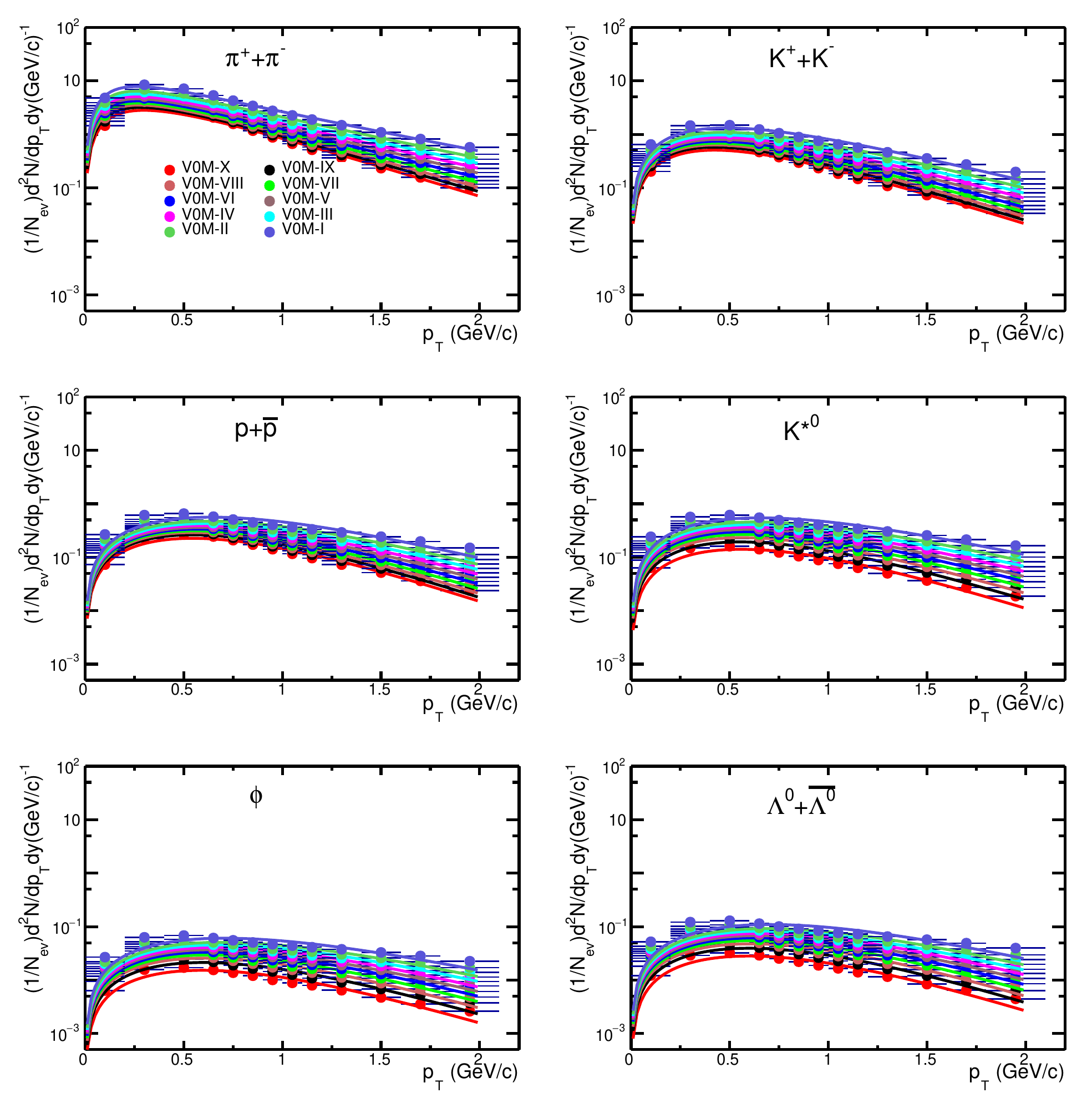}
\caption[]{(Color Online) Fitting of generated $p_{\rm{T}}$-spectra of identified hadrons from PYTHIA8 using BGBW model for spherocity integrated events in various multiplicity classes as shown in Table~\ref{tab:V0M}.}
\label{bgbw_S0}
\end{figure}

\begin{figure}[ht!]
\includegraphics[width=9cm, height=11.cm]{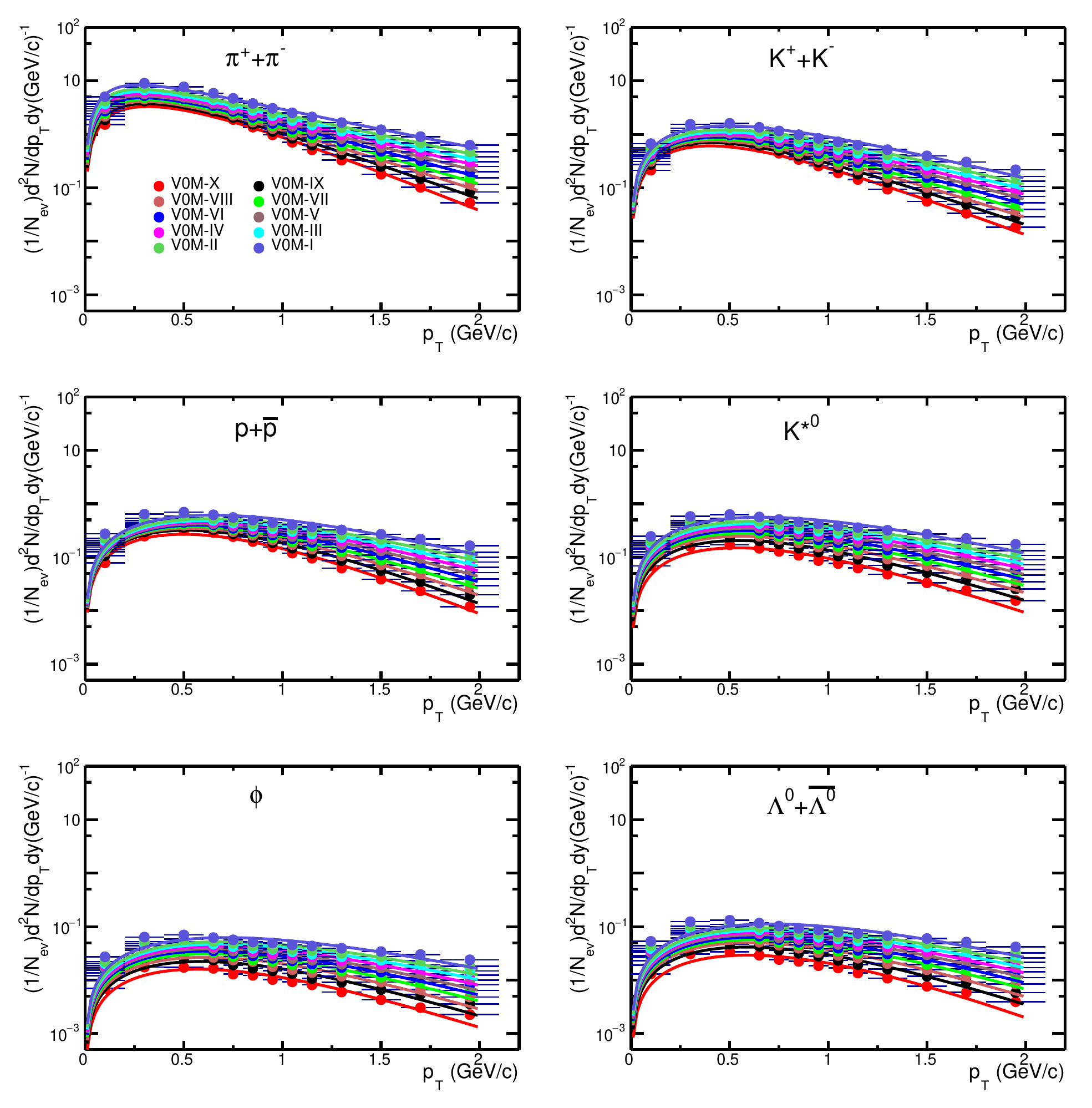}
\caption[]{(Color Online) Fitting of generated $p_{\rm{T}}$-spectra of identified hadrons from PYTHIA8 using BGBW model for isotropic events in various multiplicity classes as shown in Table~\ref{tab:V0M}.}
\label{bgbw_iso}
\end{figure}

\begin{figure}[ht!]
\includegraphics[width=9cm, height=11.cm]{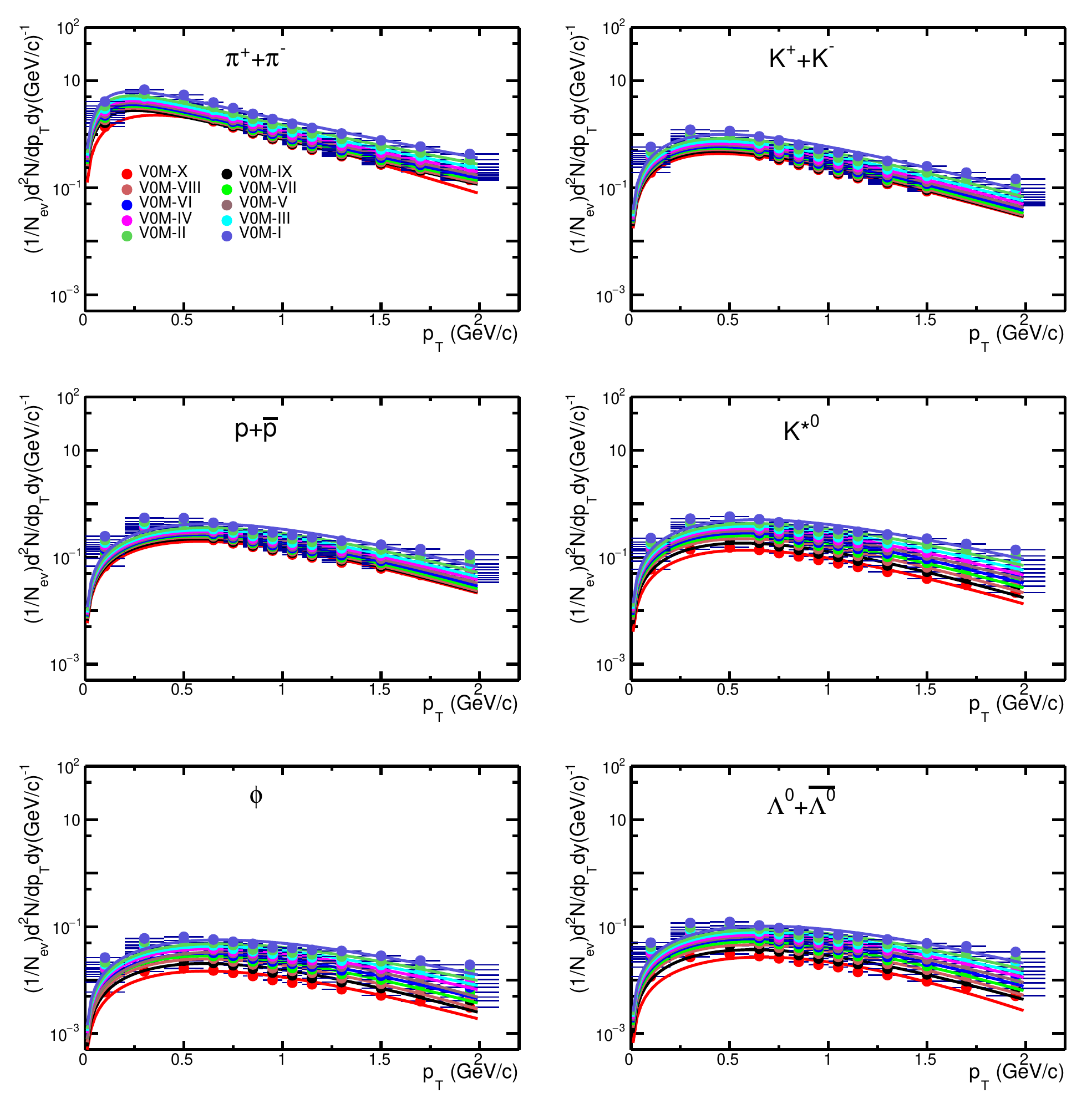}
\caption[]{(Color Online) Fitting of generated $p_{\rm{T}}$-spectra of identified hadrons from PYTHIA8 using BGBW model for jetty events in various multiplicity classes as shown in Table~\ref{tab:V0M}.}
\label{bgbw_jetty}
\end{figure}

The expression for invariant yield in Boltzmann-Gibbs Blast Wave (BGBW) model is given by~\cite{Cooper,Schnedermann:1993ws}: 
\ba
\label{eq5}
E\frac{d^3N}{dp^3}=D \int d^3\sigma_\mu p^\mu exp(-\frac{p^\mu u_\mu}{T})\,.
\ea
Here, the four-velocity denoting flow velocities in space-time is given by, $u^\mu = \cosh\rho~(\cosh\eta, \tanh\rho \cos\phi_r, \tanh\rho \sin \phi_r, \sinh \eta)$ and the particle four-momentum is, $p^\mu = (m_T\cosh y, p_T\cos\phi, p_T\sin\phi, m_T\sinh y)$, while the kinetic freeze-out surface is given by $d^3\sigma_\mu = (\cosh\eta, 0, 0, -\sinh\eta)\tau rdrd\eta d\phi_r$. Here, $\eta$ is the space-time rapidity and assuming Bjorken correlation in rapidity, $i.e.$ $y=\eta$~\cite{Bjorken:1982qr}, Eq.~\ref{eq5} can be expressed as:  
\ba
\label{eq6}
\left.\frac{d^2N}{dp_Tdy}\right|_{y=0} = D \int_0^{R_{0}} r\;dr\;K_1\Big(\frac{m_T\;\cosh\rho}{T}\Big)I_0\Big(\frac{p_T\;\sinh\rho}{T}\Big)
\ea
Here, $D = \displaystyle \frac{gVm_T}{2\pi^2}$, where $g$ is the degeneracy factor, $V$ is the system volume, and $m_{\rm T}=\sqrt{p_T^2+m^2}$ is the transverse mass. Here, $I_0\displaystyle\Big(\frac{p_T\;{\sinh}\rho}{T}\Big)$ and $K_{1}\displaystyle\Big(\frac{m_T\;{\cosh}\rho}{T}\Big)$ are the modified Bessel's functions. They are given by
\ba
\centering
K_1\Big(\frac{m_T\;{\cosh}\rho}{T}\Big)=\int_0^{\infty} {\cosh}y\;{\exp}\Big(-\frac{m_T\;{\cosh}y\;{\cosh}\rho}{T}\Big)dy,
\ea
\ba
\centering
I_0\Big(\frac{p_T\;{\sinh}\rho}{T}\Big)=\frac{1}{2\pi}\int_0^{2\pi} exp\Big(\frac{p_T\;{\sinh}\rho\;{\cos}\phi}{T}\Big)d\phi.
\ea
\begin{figure*}[ht!]
\includegraphics[width=16cm, height=10.cm]{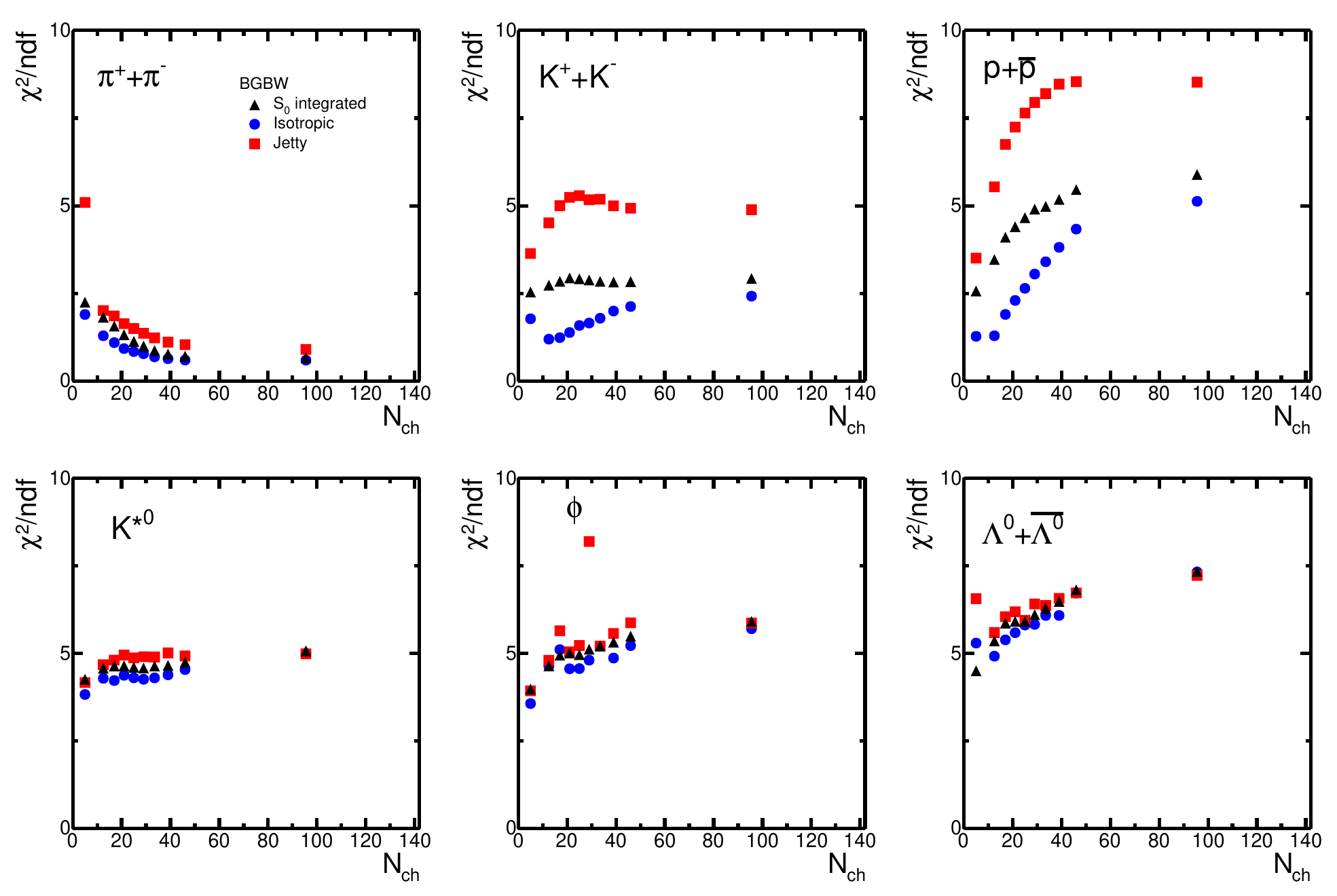}
\caption[]{(Color Online) $\chi^{2}$/NDF for the fitting of generated $p_{\rm{T}}$-spectra of identified hadrons using BGBW model in different spherocity and multiplicity classes.}
\label{bgbw_chi2}
\end{figure*}

\begin{figure*}[ht!]
\includegraphics[width=16cm, height=10.cm]{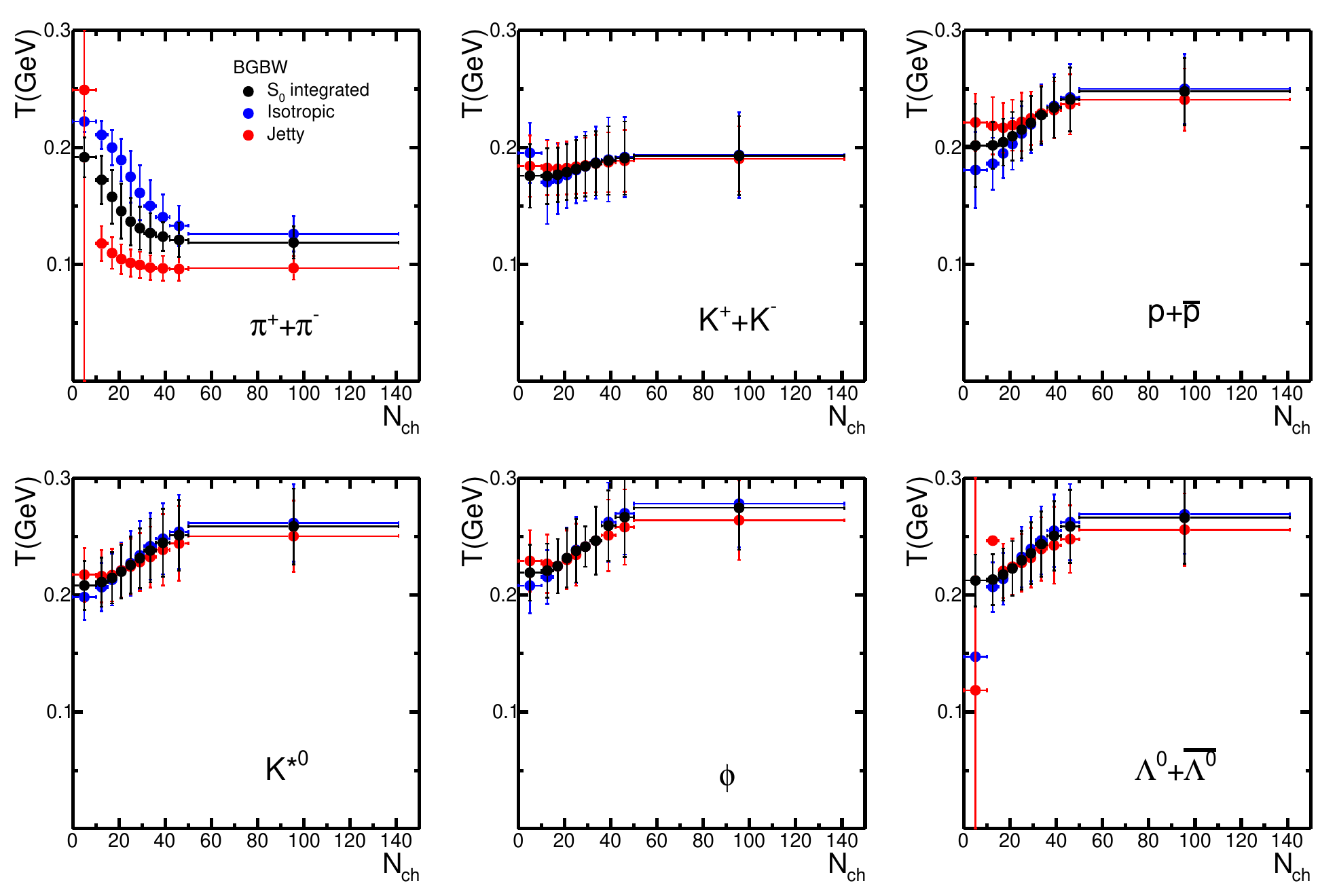}
\caption[]{(Color Online) Multiplicity dependence of $T$ in different spherocity classes from the fitting of BGBW model.}
\label{bgbw_param_T}
\end{figure*}

\begin{figure*}[ht!]
\includegraphics[width=16cm, height=10.cm]{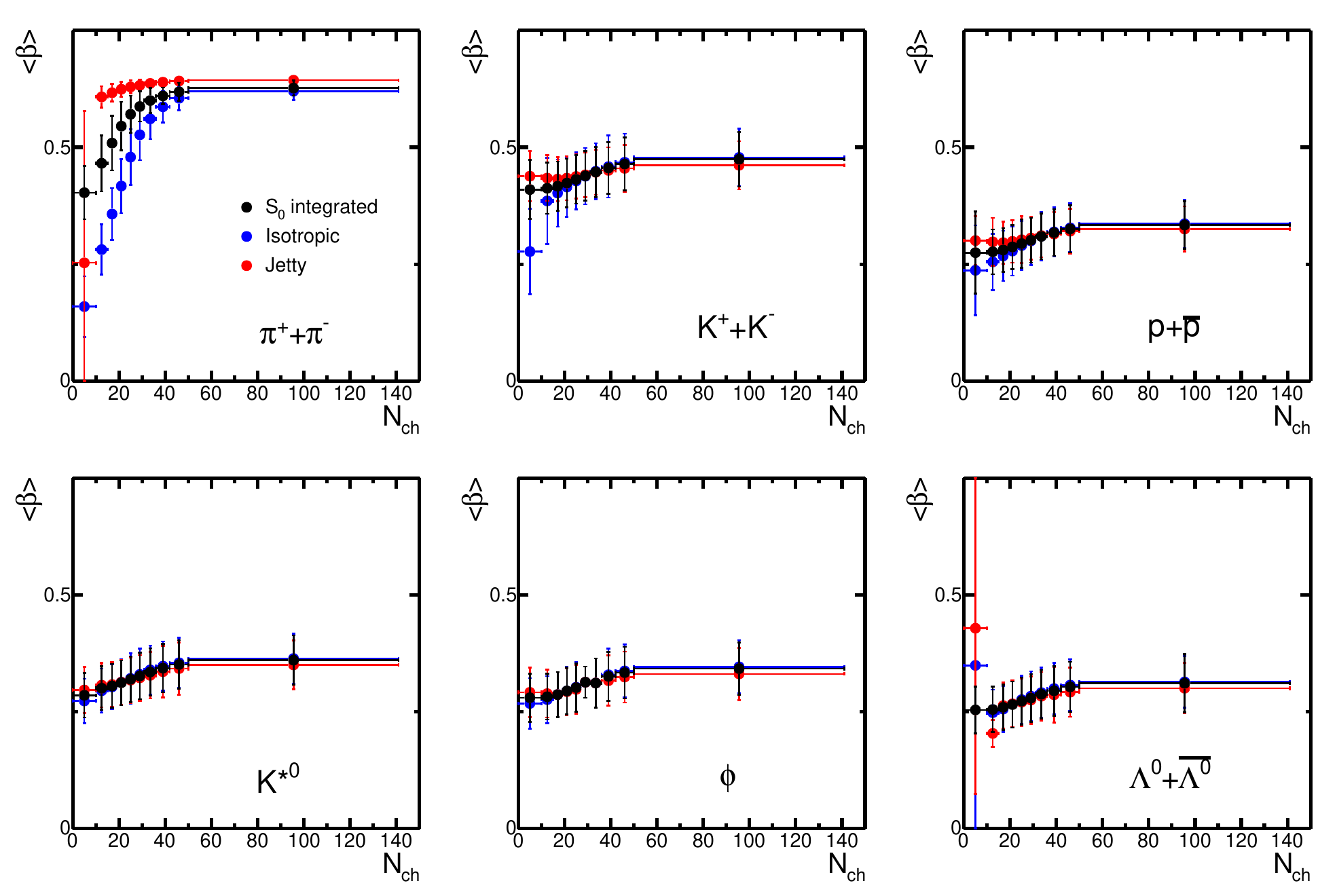}
\caption[]{(Color Online) Multiplicity dependence of $<\beta>$ in different spherocity classes from the fitting of BGBW model.}
\label{bgbw_param_beta}
\end{figure*}

Here, $\rho$ is a parameter given by $\rho=tanh^{-1}\beta$. $\beta =\displaystyle\beta_{max}\;\Big(\xi\Big)^n$ \cite{Huovinen:2001cy,Schnedermann:1993ws,BraunMunzinger:1994xr, Tang:2011xq} is the radial flow, where $\beta_{max}$ is the maximum surface velocity and $\xi=\displaystyle\Big(r/R_0\Big)$ with $r$ as the radial distance. In the BGBW model, the particles closer to the center of the fireball move slower than the ones at the edges and the average of the transverse velocity can be evaluated as \cite{Adcox:2003nr}, 
\ba
<\beta> =\frac{\int \beta_{max}\xi^n\xi\;d\xi}{\int \xi\;d\xi}=\Big(\frac{2}{2+n}\Big)\beta_{max}.
\ea
In our calculation, for the sake of simplicity we use a linear velocity profile i.e. $n=1$ and $R_0$ is the maximum radius of the expanding source at freeze-out ($0<\xi<1$).

Figures~\ref{bgbw_S0},~\ref{bgbw_iso} and~\ref{bgbw_jetty} show the fitting of $p_{\rm{T}}$-spectra of pions, kaons, protons, $\rm{K}^{*0}$, $\phi$ and $\Lambda$ as a function of charged-particle multiplicity using BGBW distribution using Eq.~\ref{eq6} for spherocity-integrated, isotropic and jetty events, respectively. The BGBW distribution fits the spectra for identified hadrons till $p_{\rm{T}}\simeq$ 2 GeV/c. Figure~\ref{bgbw_chi2} shows the $\chi^{2}$/NDF for the fitting of generated $p_{\rm{T}}$-spectra of identified hadrons using BGBW model in different spherocity and multiplicity classes. For pions, the $\chi^{2}$/NDF is relatively lower compared to kaons and protons indicating pions are better described by the BGBW model. The decrease of $\chi^{2}$/NDF for pions with increasing charged-particle multiplicity is due to the fact that the number of particles is less in the lower multiplicity classes, which makes the fitting worse. As expected, the fitting for jetty events are worse compared to isotropic and spherocity-integrated events indicating that the jetty events remain far from equilibrium and a BGBW description, hence becomes less
significant.

\begin{figure}[ht!]
\includegraphics[width=8cm, height=12.cm]{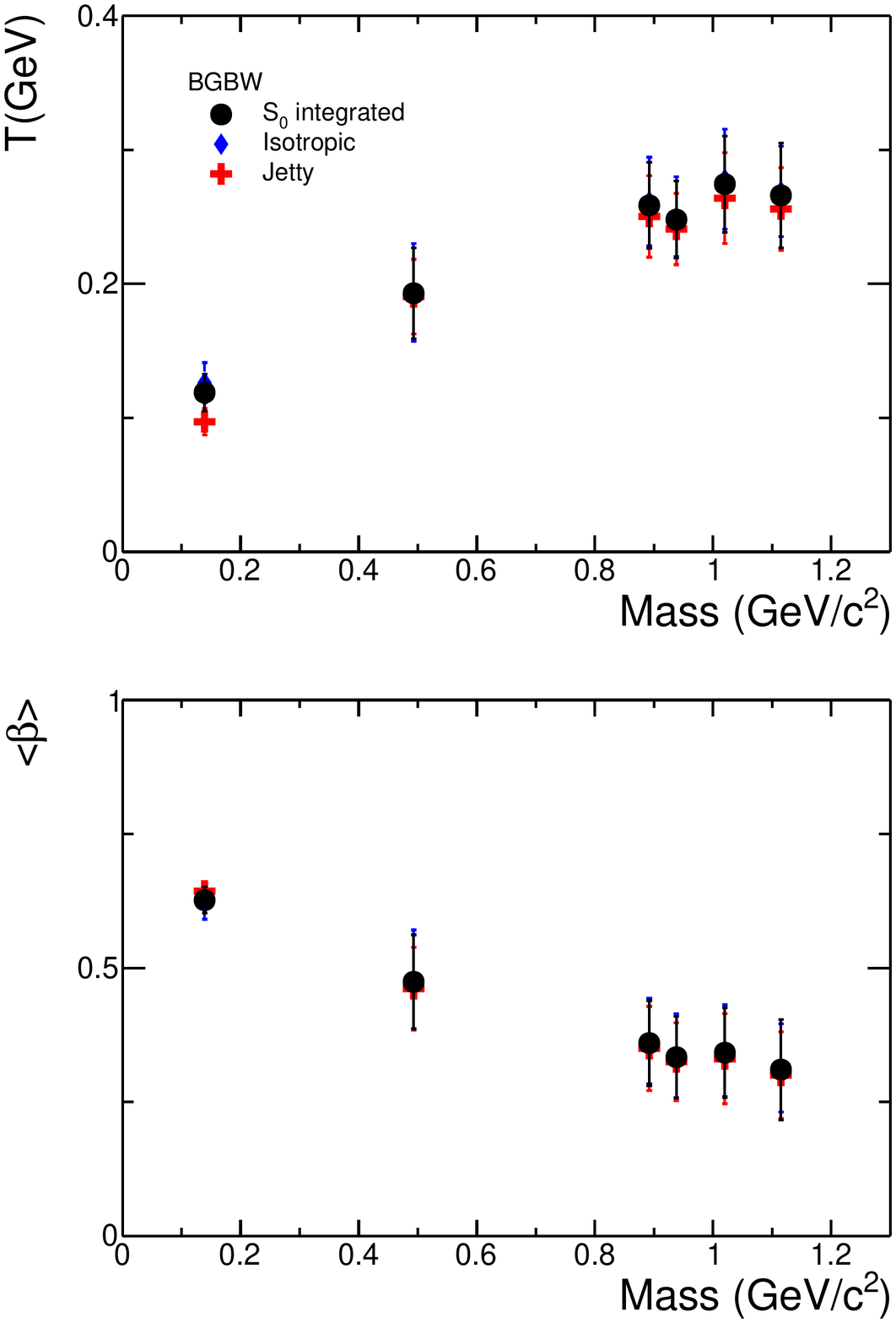}
\caption[]{(Color Online) Mass dependence of T and $<\beta>$ in different spherocity classes for the highest multiplicity class using BGBW fit up to $p_{\rm T} \sim$  2 GeV/c.}
\label{BGBW-param_mass}
\end{figure}

It is interesting to note that since BGBW analysis is in the soft sector of particle production, as expected, we 
do not see any difference between jetty, isotropic and spherocity integrated events so far the multiplicity
dependence of kinetic freeze-out  temperature and the radial flow velocity are concerned, except pions. This is
depicted in Figs.~\ref{bgbw_param_T} and \ref{bgbw_param_beta}. For all the discussed particles except the
pions, the kinetic freeze-out temperature shows a linear increase with final state multiplicity. The radial flow
velocity also shows a monotonic increase with multiplicity class for all the particles. Taking the highest multiplicity
class, let's now look into the particle mass dependence of the freeze-out parameters. Figure~\ref{BGBW-param_mass} shows that the temperature increases with mass for the highest multiplicity pp collisions indicating a differential freeze-out scenario. As seen in the previous sub-section, the temperature from BGBW model also suggests that the particles with heavier mass freeze-out early in time. The radial flow velocity is seen to decrease with particle mass, which supports a hydrodynamical behavior. We observe $<\beta>~ \simeq$  0.62 for pions, whereas it is 0.31 for $\Lambda$.

\section{Summary and Conclusion}
\label{summary}
We perform a double differential study of the identified particle spectra and the system thermodynamics as a function of charged-particle multiplicity and transverse spherocity in pp collisions at $\sqrt{s}$ = 13 TeV using PYTHIA8. 
In order to understand the production dynamics of particles in high-multiplicity pp collisions, an event shape 
dependent study becomes inevitable. Further to study the event topology dependence of the kinetic freeze-out properties, we have
taken a cosmological expansion scenario of the produced fireball with a differential particle freeze-out.
This work would shed light into the underlying event dynamics and help in understanding the possible 
differences and/or similarities in freeze-out parameters, when the hadronic collisions are compared with heavy-ion collisions. For the analysis of the identified particle $p_{\rm T}$-spectra as a function of charged-particle multiplicity and transverse spherocity, we use the thermodynamically consistent and experimentally motivated Tsallis non-extensive distribution function. In the soft sector of particle production, which corresponds to low-$p_{\rm T}$,
Boltzmann-Gibbs Blastwave (BGBW) model is used to extract the kinetic freeze-out temperature and the radial
flow velocity to study the particle mass and event multiplicity dependence. The important findings of this work
are summarized below:

\begin{itemize}
\item We observe that the temperature parameter obtained by fitting the full range of the $p_{\rm T}$-spectra using
Tsallis distribution function, is dependent on spherocity class and it increases with multiplicity for isotropic events, showing a steeper increase for higher mass particles.\\

\item The entropic parameter, $q$ is found to be spherocity and multiplicity dependent. The jetty events have a
tendency of staying away from equilibrium. For isotropic events, the $q$ values remain lower compared to the spherocity-integrated events which suggests that isotropic events approach more towards equilibrium compared to spherocity-integrated events. This hints for separating isotropic events from spherocity-integrated ones
while studying the QGP-like conditions in small systems. This is because the production dynamics are different.
In addition, while taking pp collisions as the baseline measurement to study any possible system formation in 
heavy-ion collisions at the LHC energies, the technique of transverse spherocity would be very useful.\\

\item From BGBW analysis, it is observed that the higher mass particles show higher freeze-out temperature, which is an indication of a differential freeze-out scenario.

\item The radial flow velocity is found to be mass dependent: higher for lighter mass particles- an indication
of a hydrodynamic behavior in small systems. The obtained average flow velocities indicate a substantial 
collectivity in small systems in high multiplicity pp events at the LHC energies.\\

\item The kinetic freeze-out temperature and the radial flow velocity obtained in the BGBW framework, are
observed to be independent of spherocity class, except for pions.\\

\end{itemize}

The present study is very useful in understanding the microscopic features of degrees of equilibration and their dependencies on the number of particles in the system and on the geometrical shape of an event. In the absence of speherocity dependent experimental data, the present study should give an outlook on similarities/differences between jetty and isotropic events in LHC pp events and their multiplicity dependence. This will help in making a
proper bridge in understanding the particle production from hadronic to heavy-ion collisions.

\section*{Acknowledgements}
The authors acknowledge the financial supports  from  ALICE  Project  No. SR/MF/PS-01/2014-IITI(G)  of  
Department  of  Science  \&  Technology,  Government of India. ST acknowledges the financial support by 
DST-INSPIRE program of Government of India. RS acknowledges the financial supports from DAE-BRNS Project 
No. 58/14/29/2019-BRNS of Government of India. Fruitful discussions with Suman Deb and Rutuparna Rath is highly appreciated.
The authors would like to acknowledge the usage of resources  of the LHC grid computing facility at VECC, Kolkata. 

%\section*{Declaration} This paper appears in the arXiv with No: [arXiv:1905.07418 [hep-ph]] and could be found at: https://arxiv.org/pdf/1905.07418.pdf


\begin{thebibliography}{99}

\bibitem{Collins:1974ky} 
  J.~C.~Collins and M.~J.~Perry,
  %``Superdense Matter: Neutrons Or Asymptotically Free Quarks?,''
  Phys.\ Rev.\ Lett.\  {\bf 34}, 1353 (1975).
  
  \bibitem{Cabibbo:1975ig} 
  N.~Cabibbo and G.~Parisi,
  %``Exponential Hadronic Spectrum and Quark Liberation,''
  Phys.\ Lett.\  {\bf 59B}, 67 (1975).

\bibitem{ALICE:2017jyt} 
  J.~Adam {\it et al.} [ALICE Collaboration],
  %``Enhanced production of multi-strange hadrons in high-multiplicity proton-proton collisions,''
  Nature Phys.\  {\bf 13}, 535 (2017).
  
  \bibitem{Tripathy:2018ehz} 
  S.~Tripathy [ALICE Collaboration],
  %``Energy dependence of $\phi$(1020) production at mid-rapidity in pp collisions with ALICE at the LHC,''
  Nucl.\ Phys.\ A {\bf 982}, 180 (2019).
  
  \bibitem{Dash:2018cjh} 
  A.~K.~Dash [ALICE Collaboration],
  %``Multiplicity dependence of strangeness and hadronic resonance production in pp and p-Pb collisions with ALICE at the LHC,''
  Nucl.\ Phys.\ A {\bf 982}, 467 (2019).

%\cite{Khuntia:2018znt}
\bibitem{Khuntia:2018znt} 
  A.~Khuntia, H.~Sharma, S.~Kumar Tiwari, R.~Sahoo and J.~Cleymans,
  %``Radial flow and differential freeze-out in proton-proton collisions at $\sqrt{s} = 7$ TeV at the LHC,''
  Eur.\ Phys.\ J.\ A {\bf 55}, 3 (2019).

 \bibitem{Khuntia:2018qox} 
  A.~Khuntia, S.~Tripathy, A.~Bisht and R.~Sahoo,
  %``Event Shape Engineering and Multiplicity dependent Study of Identified Particle Production in proton+proton Collisions at $\sqrt{s}$= 13 TeV using PYTHIA,''
 J.\ Phys.\ G (2021: In Press), arXiv:1811.04213 [hep-ph]. 
  
    \bibitem{Ortiz:2013yxa} 		
  A.~Ortiz Velasquez, P.~Christiansen, E.~Cuautle Flores, I.~Maldonado Cervantes and G.~Paic,
  %``Color Reconnection and Flowlike Patterns in $pp$ Collisions,''
  Phys.\ Rev.\ Lett.\  {\bf 111}, 042001 (2013).
  
 \bibitem{Khatun:2019dml}
A.~Khatun, D.~Thakur, S.~Deb and R.~Sahoo,
%``$J/\psi$  production dynamics: event shape, multiplicity and rapidity dependence in proton+proton collisions at LHC energies using PYTHIA8,''
J. Phys. G \textbf{47}, 055110 (2020).

 \bibitem{Sjostrand:2006za} 
  T.~Sjostrand, S.~Mrenna and P.~Z.~Skands,
  %``PYTHIA 6.4 Physics and Manual,''
  JHEP {\bf 0605}, 026 (2006).
  
    \bibitem{Skands:2014pea} 
  P.~Skands, S.~Carrazza and J.~Rojo, Eur.\ Phys.\ J. C {\bf 74}, 3024 (2014).
  %``Tuning PYTHIA 8.1: the Monash 2013 Tune,''
 % arXiv:1404.5630 [hep-ph].
  
   \bibitem{Adam:2015pza} 
  J.~Adam \textit{et al.} [ALICE Collaboration], 
  %``Pseudorapidity and transverse-momentum distributions of charged particles in proton–proton collisions at $\sqrt s=$ 13 TeV,'' 
  Phys. Lett. B \textbf{753}, 319 (2016).
  
    
  
   \bibitem{pythia8html}
  Pythia8 online manual:(http://home.thep.lu.se/~torbjorn/pythia81html/Welcome.html).
  
  \bibitem{Abelev:2014ffa} 
  B.~B.~Abelev {\it et al.} [ALICE Collaboration],
  %``Performance of the ALICE Experiment at the CERN LHC,''
  Int.\ J.\ Mod.\ Phys.\ A {\bf 29}, 1430044 (2014). 
  
     \bibitem{Cuautle:2014yda} 
E.~Cuautle, R.~Jimenez, I.~Maldonado, A.~Ortiz, G.~Paic and E.~Perez, 
%``Disentangling the soft and hard components of the pp collisions using the sphero(i)city approach,''
  arXiv:1404.2372 [hep-ph].
  
    \bibitem{Cuautle:2015kra} 
  A.~Ortiz, G.~Paic and E.~Cuautle,
  %``Mid-rapidity charged hadron transverse spherocity in pp collisions simulated with Pythia,''
  Nucl.\ Phys.\ A {\bf 941}, 78 (2015).
  
  \bibitem{Ortiz:2017jho}
A.~Ortiz,
%``Experimental results on event shapes at hadron colliders,''
Adv. Ser. Direct. High Energy Phys. \textbf{29}, 343 (2018).
  
      %\cite{Salam:2009jx} 
  \bibitem{Salam:2009jx} G.~P.~Salam, 
  %``Towards Jetography,''
  Eur.\ Phys.\ J.\ C {\bf 67}, 637 (2010).   
  
%    \bibitem{Bencedi:2018ctm} 
%  G.~Bencedi [ALICE Collaboration],
%  %``Event-shape- and multiplicity-dependent identified particle production in pp collisions at 13 TeV with ALICE at the LHC,''
%  arXiv:1807.09508 [hep-ex].
  
  \bibitem{Bencedi:2018ctm} 
  G.~Bencédi [ALICE Collaboration],
  %``Event-shape- and multiplicity-dependent identified particle production in pp collisions at 13 TeV with ALICE at the LHC,''
  Nucl.\ Phys.\ A {\bf 982}, 507 (2019)
  
 %\cite{Cleymans:2011in}
\bibitem{Cleymans:2011in} 
  J.~Cleymans and D.~Worku,
  %``The Tsallis Distribution in Proton-Proton Collisions at $\sqrt{s}$ = 0.9 TeV at the LHC,''
  J.\ Phys.\ G {\bf 39}, 025006 (2012).
  
   \bibitem{Cooper} 
  F. Cooper and G. Frye, Phys.\ Rev.\ D {\bf 10}, 186 (1974).
  
  %\cite{Schnedermann:1993ws}
\bibitem{Schnedermann:1993ws} 
  E.~Schnedermann, J.~Sollfrank and U.~W.~Heinz,
  %``Thermal phenomenology of hadrons from 200-A/GeV S+S collisions,''
  Phys.\ Rev.\ C {\bf 48}, 2462 (1993).

 \bibitem{Hagedorn:1965st} 
  R.~Hagedorn,
  %``Statistical thermodynamics of strong interactions at high-energies,''
  Nuovo Cim.\ Suppl.\  {\bf 3}, 147 (1965).
  
 \bibitem{CM} C. Michael and L. Vanryckeghem,  J.\ Phys.\ G {\bf 3} L151 (1977).
 
 \bibitem{CM1} C. Michael,  Prog.\ Part.\ Nucl.\ Phys. {\bf 2}, 1 (1979).
 
 \bibitem{UA1} G. Arnison {\it et al.} (UA1 Collaboration),  Phys.\ Lett.\ B {\bf 118}, 167 (1982).
 
 \bibitem{Hagedorn:1983wk} 
  R.~Hagedorn,
  %``Multiplicities, $p_{\rm T}$ Distributions and the Expected Hadron $\to$ Quark - Gluon Phase Transition,''
  Riv.\ Nuovo Cim.\  {\bf 6N10}, 1 (1983).
  
\bibitem{star-prc75} B. I. Abelev {\it et al.} (STAR Collaboration), Phys.\ Rev.\ C {\bf 75}, 064901 (2007).

\bibitem{phenix-prc83} A. Adare {\it et al.} (PHENIX Collaboration), Phys.\ Rev.\ C {\bf 83}, 064903 (2011).

\bibitem{alice1} K. Aamodt {\it et al.} (ALICE Collaboration), Eur.\ Phys.\ J.\ C {\bf 71}, 1655 (2011).

\bibitem{alice2} B. Abelev {\it et al.} (ALICE Collaboration), Phys.\ Letts.\  B {\bf 717}, 162 (2012).

\bibitem{alice3} B. Abelev {\it et al.} (ALICE Collaboration), Phys.\ Letts.\  B {\bf 712}, 309 (2012).

\bibitem{cms} S. Chatrchyan {\it et al.} (ALICE Collaboration), Eur.\ Phys.\ J.\ C {\bf 72}, 2164 (2012).

 \bibitem{Bhattacharyya:2015nwa} 
  T.~Bhattacharyya, P.~Garg, R.~Sahoo and P.~Samantray,
  %``Time Evolution of Temperature Fluctuation in a Non-Equilibrated System,''
  Eur.\ Phys.\ J.\ A {\bf 52}, 283 (2016).
  
   \bibitem{Bhattacharyya:2015hya} 
  T.~Bhattacharyya, J.~Cleymans, A.~Khuntia, P.~Pareek and R.~Sahoo,
  %``Radial Flow in Non-Extensive Thermodynamics and Study of Particle Spectra at LHC in the Limit of Small $(q-1)$,''
  Eur.\ Phys.\ J.\ A {\bf 52}, 30 (2016).
  
  \bibitem{Zheng:2015gaa} 
  H.~Zheng and L.~Zhu,
  %``Can Tsallis Distribution Fit All the Particle Spectra Produced at RHIC and LHC?,''
  Adv.\ High Energy Phys.\  {\bf 2015}, 180491 (2015).
  
\bibitem{Tang:2008ud} Z.~Tang, Y.~Xu, L.~Ruan, G.~van Buren, F.~Wang and Z.~Xu,
  %``Spectra and radial flow at RHIC with Tsallis statistics in a Blast-Wave description,''
  Phys.\ Rev.\ C {\bf 79}, 051901 (2009).
  
\bibitem{De:2014dna} B.~De,
  %``Non-extensive statistics and understanding particle production and kinetic freeze-out process from p$_{T}$-spectra at 2.76 TeV,''
  Eur.\ Phys.\ J.\ A {\bf 50}, 138 (2014).  
   
   \bibitem{e+e-} I. Bediaga, E.M.F. Curado, J.M. de Miranda, Physica A {\textbf{286}}, 156 (2000).
   
\bibitem{R1} G. Wilk and Z. W\l{}odarczyk, Acta Phys. Polon. B {\textbf{46}}, 1103 (2015).

\bibitem{R2} K. \"Urm\"ossy, G.G. Barnaf\"{o}ldi, T.S. Bir\'{o}, Phys. Lett. B {\textbf{701}}, 111 (2011).

\bibitem{R3} K. \"Urm\"ossy, G.G. Barnaf\"{o}ldi, T.S. Bir\'{o}, Phys. Lett. B {\textbf{718}}, 125 (2012).

\bibitem{ijmpa} P. K. Khandai, P. Sett, P. Shukla, V. Singh, Int. Jour. Mod. Phys. A {\textbf{28}}, 1350066 (2013).

\bibitem{plbwilk} B.-C. Li, Y.-Z. Wang and F.-H. Liu, Phys. Lett. B {\textbf{725}}, 352 (2013).

\bibitem{marques} L. Marques, J. Cleymans and A. Deppman Phys. Rev. D{\textbf{91}}, 054025 (2015).

  \bibitem{STAR} B. I. Abelev et al. (STAR  collaboration), Phys. Rev. C {\textbf{75}}, 064901 (2007).
  
\bibitem{PHENIX1} A. Adare et al. (PHENIX collaboration), Phys. Rev. D {\textbf{83}}, 052004 (2011).
 
\bibitem{PHENIX2} A. Adare et al. (PHENIX collaboration), Phys. Rev. C {\textbf{83}}, 064903 (2011).

\bibitem{ALICE_charged} K. Aamodt, et al. (ALICE collaboration), Phys. Lett. B {\textbf{693}}, 53 (2010).

\bibitem{ALICE_piplus} K. Aamodt, et al. (ALICE collaboration), Eur. Phys. J C {\textbf{71}}, 1655 (2011).

\bibitem{CMS1} V. Khachatryan, et al. (CMS collaboration), J. of High Eng. Phys. {\textbf{02}}, 041 (2010).

\bibitem{CMS2} V. Khachatryan, et al. (CMS collaboration), Phys. Rev. Lett. {\textbf{105}}, 022002 (2010).

\bibitem{ATLAS} G. Aad, et al. (ATLAS collaboration), New J. Phys. {\textbf{13}}, 053033 (2011).
%\bibitem{ALICE_PbPb} K. Aamodt, et al. (ALICE collaboration), Eur. Phys. J. C {\textbf{73}} (2013) 2662.

\bibitem{ALICE_PbPb} 
B. Abelev,  et al. (ALICE collaboration), Phys. Rev. Letts. {\textbf{109}},  252301 (2012).

\bibitem{Grigoryan:2017gcg} 
  S.~Grigoryan,
  %``Using the Tsallis distribution for hadron spectra in $pp$ collisions: Pions and quarkonia at $\sqrt{s}$ = 5--13000 GeV,''
  Phys.\ Rev.\ D {\bf 95}, 056021 (2017).
  
  \bibitem{Parvan:2016rln} 
  A.~S.~Parvan, O.~V.~Teryaev and J.~Cleymans,
  %``Systematic Comparison of Tsallis Statistics for Charged Pions Produced In $pp$ Collisions,''
  Eur.\ Phys.\ J.\ A {\bf 53}, 102 (2017).
  
   \bibitem{Cleymans:2013rfq} J.~Cleymans, G.~I.~Lykasov, A.~S.~Parvan, A.~S.~Sorin, O.~V.~Teryaev and D.~Worku, %``Systematic properties of the Tsallis Distribution: Energy Dependence of Parameters in High-Energy p-p Collisions,''
  Phys.\ Lett.\ B {\bf 723}, 351 (2013).
  
  \bibitem{book}  C.~Tsallis,
 Introduction to Nonextensive Statistical  Mechanics (Springer,  2009).
  
   \bibitem{Beck:2003kz} C.~Beck, %``Superstatistics: Theory and applications,''
  Contin.\ Mech.\ Thermodyn.\ {\bf 16}, 293 (2004).
  
  \bibitem{Wilk} G.~Wilk and Z.~Wlodarczyk, %``On the interpretation of nonextensive parameter q in Tsallis statistics and Levy distributions,''
  Phys.\ Rev.\ Lett.\ {\bf 84}, 2770 (2000).
 
 %\cite{Bjorken:1982qr}
\bibitem{Bjorken:1982qr} 
  J.~D.~Bjorken,
  %``Highly Relativistic Nucleus-Nucleus Collisions: The Central Rapidity Region,''
  Phys.\ Rev.\ D {\bf 27}, 140 (1983).
  
  \bibitem{Huovinen:2001cy} 
  P.~Huovinen, P.~F.~Kolb, U.~W.~Heinz, P.~V.~Ruuskanen and S.~A.~Voloshin,
  %``Radial and elliptic flow at RHIC: Further predictions,''
  Phys.\ Lett.\ B {\bf 503}, 58 (2001).

    %\cite{BraunMunzinger:1994xr}
\bibitem{BraunMunzinger:1994xr} 
  P.~Braun-Munzinger, J.~Stachel, J.~P.~Wessels and N.~Xu,
  %``Thermal equilibration and expansion in nucleus-nucleus collisions at the AGS,''
  Phys.\ Lett.\ B {\bf 344}, 43 (1995).
  
  %\cite{Tang:2011xq}
\bibitem{Tang:2011xq}
  Z.~Tang {\it et al.},
  %``Statistical Origin of Constituent-Quark Scaling in the QGP hadronization,''
  Chin.\ Phys.\ Lett.\  {\bf 30}, 031201 (2013).
  
    
  \bibitem{Adcox:2003nr} 
  K.~Adcox {\it et al.} [PHENIX Collaboration],
  %``Single identified hadron spectra from s(NN)**(1/2) = 130-GeV Au+Au collisions,''
  Phys.\ Rev.\ C {\bf 69}, 024904 (2004). 

  \end{thebibliography}
  \end{document}